\documentclass[a4paper,11pt]{article}
\pdfoutput=1
\usepackage[a4paper,
  left=2.5cm, right=2.5cm,
  top= 3cm, bottom=4cm]{geometry}

\usepackage{amsmath}
\numberwithin{equation}{section}
\usepackage{mathtools}
\usepackage{oldgerm}
\usepackage{amssymb}
\usepackage{bbm}
\usepackage{graphicx}
\usepackage{subcaption}
\usepackage{tikz}
\usepackage{tikz-cd} 
\usetikzlibrary{backgrounds, arrows,calc,shapes,decorations.pathreplacing, decorations.markings, automata,positioning}
\usepackage{standalone}
\usepackage{epic}
\usepackage{bbold}

\newcommand{\be}{\begin{equation}}  
\newcommand{\ee}{\end{equation}}  
\newcommand{\bp}{\begin{pmatrix*}[r]}  
\newcommand{\ep}{\end{pmatrix*}}  
\newcommand{\bpp}{\begin{pmatrix}}  
\newcommand{\epp}{\end{pmatrix}}  
\newcommand{\bcd}{\begin{equation}
\begin{tikzcd}}
\newcommand{\ecd}{\end{tikzcd} \end{equation}}

\def\P{\mathbb{P}}
\def\cL{\mathcal{L}}
\def\cC{\mathcal{C}}
\def\C{\mathbb{C}}
\def\cO{\mathcal{O}}
\def\cS{\mathcal{S}}
\def\1{\mathbb{1}}

\usepackage{hyperref}
\begin{document}
\begin{titlepage}
 
\vskip -0.5cm
\rightline{\small{\tt  t14/167}} 
 
\begin{flushright}

\end{flushright}
 
\vskip 1cm
\begin{center}
 
{\huge \bf \boldmath T-branes as branes within branes} 
 
 \vskip 2cm
 
Andr\'es Collinucci$^1$ and Raffaele Savelli$^{2}$

 \vskip 0.4cm
 
 {\it  $^1$Physique Th\'eorique et Math\'ematique and International Solvay Institutes,\\ Universit\'e Libre de Bruxelles, C.P. 231, 1050
Bruxelles, Belgium \\[2mm]
 
 $^2$Institut de Physique Th\'eorique, CEA Saclay, Orme de Merisiers, F-91191 Gif-surYvette, France
 }
 \vskip 2cm
 
\abstract{Bound states of 7-branes known as `T-branes' have properties that defy usual geometric intuition. For instance, the gauge group of $n$ coincident branes may not be $U(n)$. More surprisingly, matter may show up at unexpected loci, such as points.

By analyzing T-branes of perturbative type IIB string theory in the tachyon condensation picture we gain the following insights: In a large class of models, the tachyon can be diagonalized even though the worldvolume Higgs cannot. In those cases, we see the structure of these bound states more manifestly, thereby drastically simplifying analysis of gauge groups and spectra.
Whenever the tachyon is not diagonalizable, matter localizes at unexpected loci, and we find that there is a lower-dimensional brane bound to the 7-brane.} 

\end{center}

\end{titlepage}

\tableofcontents

\section{Introduction}

Conventional wisdom on type IIB compactifications with 7-branes offers a nice geometric hierarchy for the various ingredients that make up the four-dimensional effective theory: Gravity propagates along all of spacetime, gauge fields are localized on divisors of the internal manifold, matter fields are localized on Riemann surfaces, and Yukawa couplings are point-like.

However, this geometric intuition is amenable to drastic modifications once one allows worldvolume scalars to acquire non-trivial vev's. For instance, on a stack of $n$ D-branes we expect to see the spectrum of a $U(n)$ gluon. On D7-branes, the worldvolume theory also carries an adjoint-valued complex scalar. If this scalar acquires a vev along a Cartan generator, it comes as no surprise that the gauge group gets broken to some subgroup. E.g. 

\be 
\langle \Phi \rangle = {\rm diag}(\phi_1, \phi_2, \ldots, \phi_n) \qquad U(n) \rightarrow U(1)^n\,.
\ee
After all, such a vev has the geometric interpretation of separating the branes onto the positions $z=\phi_i$, for the complex transverse coordinate $z$, thereby making part of the spectrum massive.

However, in \cite{Donagi:2003hh}, new vev's were explored that do not alter the geometry of the stack of D-branes, yet they drastically alter the gauge group. These bound states were later analyzed in more generality in \cite{Donagi:2011jy} and \cite{Cecotti:2010bp}. In the latter, they were dubbed `T-branes', and were later analyzed in the F-theory context in \cite{Anderson:2013rka}. The `T' stands for `triangular', meaning the Higgs can be given a nilpotent, strictly triangular vev. For example, the vev 
\be
\Phi = \bp 0 & 1 \\ 0 & 0 \ep
\ee
breaks $U(2)$ to the overall decoupled $U(1)$.

General 7-brane vacua can host a number of other surprising effects. Several were listed in section 7.2 of \cite{Cecotti:2010bp}. The most surprising is perhaps the fact that charged matter can sometimes be absent at the intersection of two branes, or conversely, there can be matter localized along a Riemann surface that has nothing to do with brane intersections. Finally, matter can also localize at points.
An example of such enigmatic behavior was studied in detail in \cite{Cecotti:2010bp} from the worldvolume gauge theory point of view, but, as the authors claimed, the physical meaning remained obscure.

In this paper, we recast T-branes in Sen's tachyon condensation picture \cite{Sen:1998sm}. More precisely, we describe perturbative type IIB D7-branes in the language of the derived category of coherent sheaves \cite{Douglas:2000gi}, which is the most natural one for describing B-branes. This move drastically simplifies the analysis.

First of all, by using this picture, one can almost trivially read off the unbroken gauge group of a system by diagonalizing the tachyon field. Unlike the 7-brane worldvolume Higgs, which cannot be diagonalized when a T-brane background is switched on, the tachyon field does allow this under a wide class of circumstances.

In the cases with non-diagonalizable tachyon, we find matter fields that localize at unexpected locations, or unexpected dimensions. By using our picture, we elucidate this mysterious behavior: In all cases analyzed, the cause always turns out to be that there is a bound state with a lower-dimensional brane that is causing the localization.

We will focus on B-branes in the B-model. In other words, we will mainly set aside the issue of stability, and study only the holomorphic data. Physically speaking, this is of course an oversimplification of the models we will discuss. It will allow us to focus on what we consider to be the salient features of T-branes: Unexpected gauge symmetry breaking patterns, and unexpected localization of chiral matter. The issue of stability is of course important, as it decides whether a T-brane bound state is possible or forced upon us. In short, the holomorphic analysis we will emphasize determines what is `kinematically' possible, whereas stability determines what is `dynamically' allowed. 

This paper is organized as follows. In section \ref{TachyonCondensationReview} we review the tachyon condensation description of D7-branes and introduce the mathematical toolkit needed later. In section \ref{sec:coherentsheaves} we discuss four different instances of T-brane backgrounds in type IIB string theory, elucidating their peculiar behavior. In all cases, we compare our approach to the standard analysis based on the study of the worldvolume Higgs profile. In section \ref{TspectrumSec} we focus on general T-brane backgrounds of $U(n)$ gauge theories and study their spectrum of fluctuations. In section \ref{sec:stability} we will present some compact examples, where $\Pi$-stability is taken into account. Finally, in section \ref{DiscSec} we speculate about generalizations of our description of T-branes to non-perturbative F-theory configurations.

\section{D7-branes as D9/anti-D9 tachyon condensates}\label{TachyonCondensationReview}

It has been known for a while that all physically consistent D-branes in type IIB can be described as the by-product of tachyon condensation between D9's and anti-D9's \cite{Sen:1998sm}. This idea gave rise to the K-theory treatment of branes \cite{Freed:1999vc,Witten:1998cd,Minasian:1997mm}, and eventually to the program of using the derived category of coherent sheaves \cite{Douglas:2000gi,Sharpe:1999qz}. 
In this chapter, we will briefly review this picture (see \cite{Aspinwall:2004jr} and \cite{Sharpe:2003dr} for more detailed introductions). Introducing this formalism will payoff in two ways: It will significantly simplify the treatment of T-branes, and it will pave the way for a companion paper \cite{Collinucci:2014taa}, in which we will describe general F-theory setups through a related formalism known as \emph{matrix factorizations}.

For D7-branes, the realization is very simple. Suppose we want to describe a D7-brane located at $P=0$, where $P$ is some polynomial. Define a D9-brane with a gauge line-bundle $\cL_1$, and an anti-D9-brane with $\cL_2$. This system will be unstable, with a bifundamental tachyonic string represented by a field $T$, with
\be
T \in \Gamma(\cL_1 \otimes \cL_2^*)\,.
\ee
If this tachyon has a profile given literally by the very polynomial $\langle T \rangle = P$, then the brane/anti-brane annihilation will only be partial, leaving behind a D7-brane at $P=0$. This is summarized by the following short exact sequence:
\be
0 \longrightarrow \cL_2 \stackrel{\cdot P}\longrightarrow \cL_1 \longrightarrow \mathcal{S} \longrightarrow 0 \,.
\ee
Here, the cokernel of the map $\cdot P$ is a sheaf denoted by $\mathcal{S}$ with support only over $P=0$. Everywhere else, the map is invertible, and so the cokernel is empty.

\begin{figure}
		\begin{subfigure}[b]{0.5\textwidth} 
			\includegraphics[width=\textwidth]{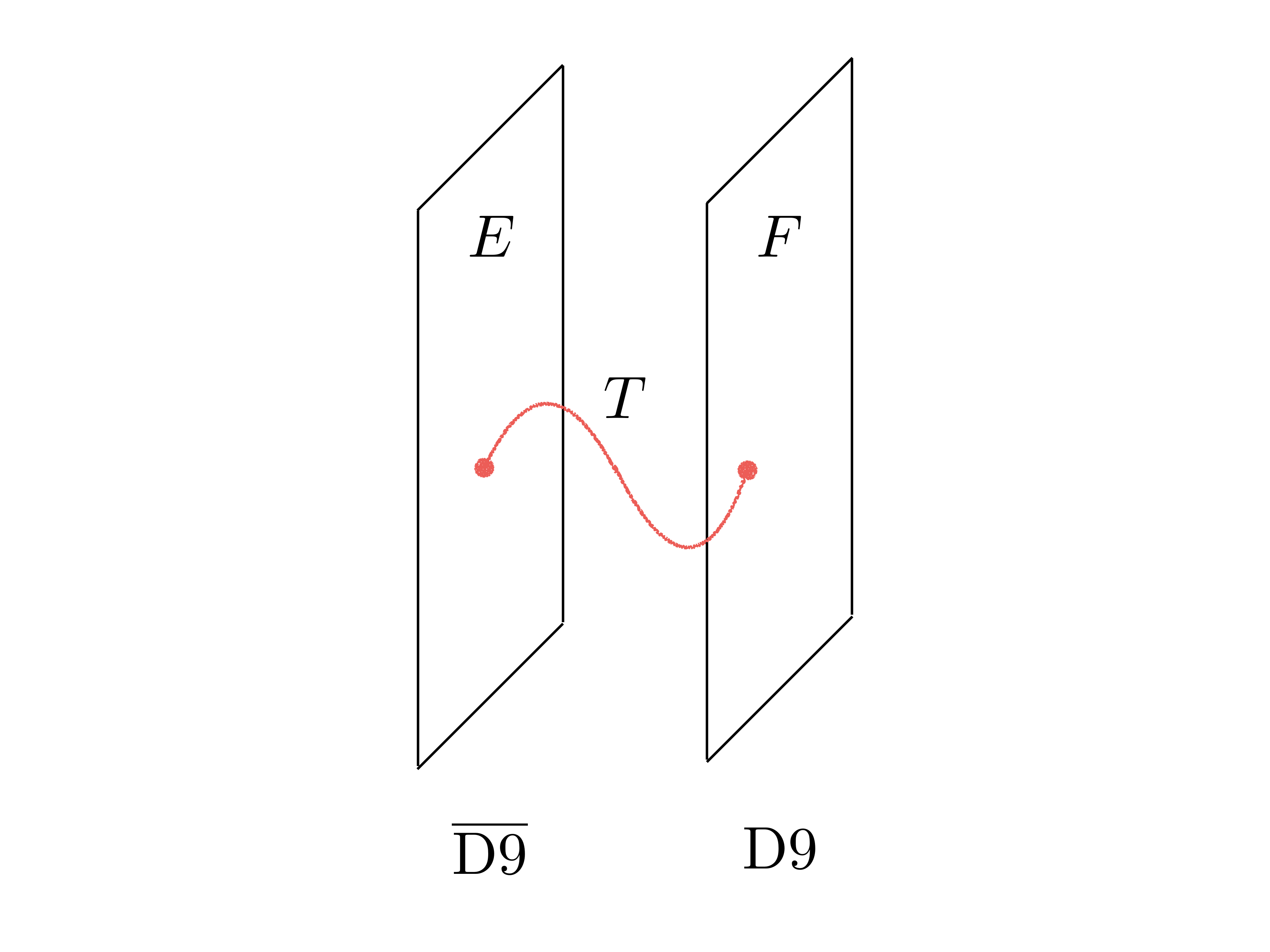}
			\caption{Two term complex} \label{tachyon2term}
		\end{subfigure}
		\begin{subfigure}[b]{0.5\textwidth}
			\includegraphics[width=\textwidth]{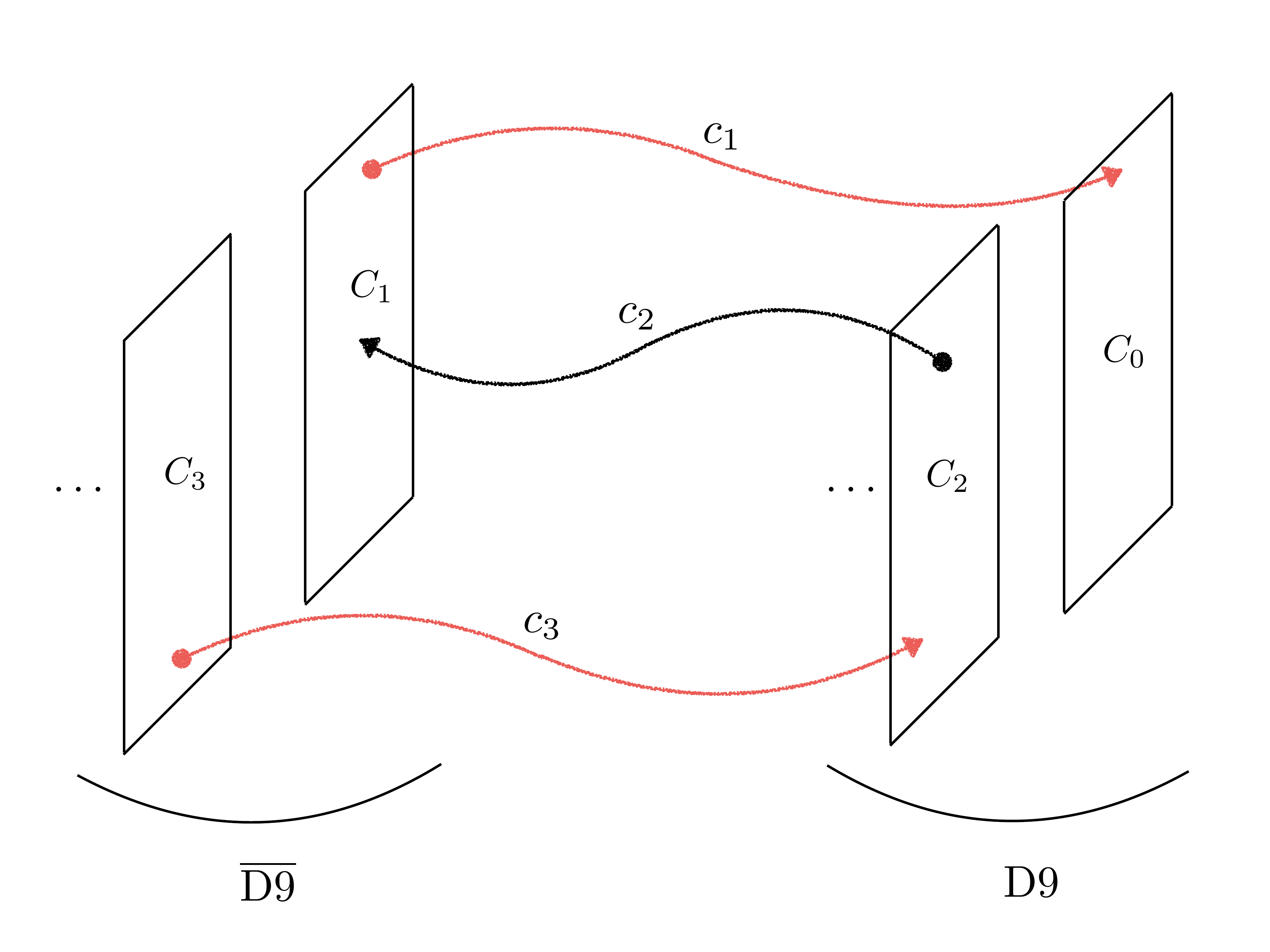}
			\caption{General complex} \label{tachyoncomplex}
		\end{subfigure}
		\caption{Figure (a) depicts a system where the tachyon field is realized by strings going from the anti-D9's to the D9's. Figure (b) depicts a more general situation with strings going back and forth between D9 and anti-D9. The data is unfolded into a complex.}
	\end{figure}

To create a non-Abelian stack of D7-branes, we start with a non-Abelian stack of D9's with gauge bundle $F$, and a stack of anti-D9's of equal rank with gauge bundle $E$, as depicted in figure \ref{tachyon2term}. The tachyonic string going from the anti-branes to the branes is a bundle map $T$:
\be
\begin{tikzcd}
E \rar{T} & F\,.
\end{tikzcd}
\ee
As a field, $T$ transforms in the bifundamental w.r.t. to the gauge groups on the D9's and anti-D9's 

\begin{equation} \label{BifundTransfT}
\begin{tikzpicture}[scale=2]
\node (E) at (0,0) {$E$};
\node[right = of E] (F)   {$F$};
\node[right = of F] (ar) {$\Longrightarrow$};
\node[right = of ar] (T) {$T$};
\node[right = of T] (Tprime) {$g_{\rm D9}\, T \, g_{\overline{\rm D9}}^{-1}$};

 \draw (E) edge[->] node[above, font=\scriptsize]{$T$} (F);
 \draw (T) edge[->] (Tprime);
 \draw(E) edge[loop below] node[font=\scriptsize]{$g_{\overline{\rm D9}}$} (E) ;
 \draw(F) edge[loop below] node[font=\scriptsize]{$g_{\rm D9}$} (F) ;
\end{tikzpicture} 
\end{equation}

If $T$ assumes a nowhere singular (i.e. everywhere invertible) profile, then there will be perfect brane/anti-brane annihilation, leaving nothing behind. However, if $T$ is a non-trivial section of $E^* \otimes F$, then the annihilation will fail whenever $T$ is non-invertible.

For instance, consider an $U(2)$ stack of D7-branes at $z=0$ in $\C \times \mathbb{R}^{1,7}$. We will suppress the irrelevant eight-dimensional factor. Let $S \equiv \C[z]$ be the polynomial ring of $\C$, which is essentially the trivial line bundle of holomorphic functions over $\C$. Then the exact sequence of interest is the following:
\be
\begin{tikzcd}
0 \arrow{r} & S^{\oplus 2} \arrow{r}{T} & 
S^{\oplus 2} \arrow{r} & \mathcal{S}_{U(2)} \rar & 0\,,
\end{tikzcd}
\ee
where $T$ is the tachyon with profile 
\be\label{su2tachyon}
T = \bp z & 0\\0&z \ep\,.
\ee
Here, the first two terms correspond to a stack of two anti-D9's, and a stack of two D9's, respectively. The last object is a sheaf corresponding to a rank-two bundle with support over $z=0$. To see the residual gauge group on the D7-brane, we start by noting that the D9-stack and the anti-D9-stack have each an $U(2)$ gauge group of their own. The tachyon field transforms in the bi-fundamental of these two $U(2)$'s, as in \eqref{BifundTransfT}.
The residual gauge group after the tachyon condenses is the one that leaves its form invariant. For the background \eqref{su2tachyon} only the diagonal $U(2)$ survives, i.e. $g_{\rm D9} = g_{\overline{\rm D9}}$.

The matter on this system is given by the fluctuations $\delta T$ of the tachyon modulo linearized gauge transformations $g = \mathbb{1}+\mathfrak{g}$:
\be \label{gaugeequivalence} \delta T \sim \delta T+\mathfrak{g}_{\rm D9} \, T- T \, \mathfrak{g}_{\overline{\rm D9}}\,.
\ee
To see this, let us make our description of branes slightly more complete. So far, we only considered a tachyonic string going from the anti-D9's to the D9's, but there can also be tachyons going in the opposite direction. It turns out that one can `unfold' this information into a complex of bundles as depicted in figure \ref{tachyoncomplex}:
\be
\begin{tikzcd}
C_\bullet: & C_n \rar{c_n} & C_{n-1} \rar{c_{n-1}} & \ldots \rar{c_1} & C_0\,.
\end{tikzcd}
\ee
Here, each $C_i$ is a spacetime filling D9-brane or anti-D9-brane if $i$ is even or odd, respectively. So, in terms of the previous description
\be
E = \bigoplus_{odd} C_i\,, \quad {\rm and} \quad F = \bigoplus_{even} C_i\,.
\ee
The maps $c_i$ are pieces of the full tachyonic spectrum of the system, organized such that they compose to zero 
\be c_{i-1} \circ c_i = 0 \quad \forall i\,. \ee
One can perform gauge transformations on each constituent brane or antibrane, i.e. automorphisms on each of the $C_i$, as long as one transforms the maps accordingly as bifundamentals
\begin{equation} 
\begin{tikzpicture}[scale=2]
\node (Cn) at (0,0) {$C_n$};
\node[right = of Cn] (Cn-1)   {$C_{n-1}$};
\node[right = of Cn-1] (dots)   {$\ldots$};
\node[right = of dots] (C0) {$C_0$};
\node[right = of C0] (with) {with};
\node[right = of with] (ci) {$c_i$};
\node[right = of ci] (ciprime) {$g_{i-1} c_i \, {g_i}^{-1}$};
\node[right = of ciprime] (forevery) {$\forall i$};

\draw (Cn) edge[->] node[auto, font=\scriptsize]{$c_n$} (Cn-1);
\draw (Cn-1) edge[->] node[auto, font=\scriptsize]{$c_{n-1}$} (dots);
\draw (dots) edge[->] node[auto, font=\scriptsize]{$c_1$} (C0);
\draw (ci) edge[->]  (ciprime);
\draw (Cn) edge[loop below] node[font=\scriptsize]{$g_n$} (Cn);
\draw (Cn-1) edge[loop below] node[font=\scriptsize]{$g_{n-1}$} (Cn-1);
\draw (C0) edge[loop below] node[font=\scriptsize]{$g_{0}$} (C0);

\end{tikzpicture} 
\end{equation}

The adjoint matter is given by the so-called Ext$^1(C_\bullet, C_\bullet)$, defined as the set of vertical maps $\{m_k\}$ in the following commutative diagram
\be
\begin{tikzcd}[row sep=large, column sep=large]
& C_n \dar{m_n} \rar{c_n} & C_{n-1} \dar{m_{n-1}} \rar{c_{n-1}} & \ldots \rar{c_2}& C_1 \dar{m_1} \rar{c_1} & C_0 \\
C_n \rar{c_n} & C_{n-1} \rar{c_{n-1}} & C_{n-2}\rar{c_{n-2}} & \ldots \rar{c_1} & C_0 
\end{tikzcd}
\ee
i.e. such that $c_{i-1} \circ m_i = m_{i-1} \circ c_i \;\;\forall i$. Such a collection of maps is referred to as a \emph{morphism} between two complexes. We actually want to mod out by morphisms that are gauge equivalent to zero. Those are morphisms equivalent to \emph{homotopies}, i.e. diagonal maps $\{d_k\}$ in 
\be
\begin{tikzcd}[row sep=huge, column sep=large]
& C_n \dlar[near start, dashed]{d_n} \dar{m_n} \rar{c_n} & C_{n-1} \dar{m_{n-1}} \dlar[near start, dashed]{d_{n-1}} \rar{c_{n-1}} & \ldots \rar{c_2} \dlar[near start, dashed]{d_{n-2}} & C_1 \dar{m_1} \dlar[near start, dashed]{d_1} \rar{c_1} & C_0 \dlar[near start, dashed]{d_0} \\
C_n \rar{c_n} & C_{n-1} \rar{c_{n-1}} & C_{n-2}\rar{c_{n-2}} & \ldots \rar{c_1} & C_0 
\end{tikzcd}
\ee
such that $m_i = c_i \circ d_i+d_{i-1} \circ c_i \; \;\forall i$.

In our running example, our brane $\cS_{U(2)}$ is represented by the two-term complex 
\be
\begin{tikzcd}
\cS_{U(2)}: & S^{\oplus 2} \arrow{r}{T} & S^{\oplus 2}\,.
\end{tikzcd}
\ee
Therefore, the adjoint matter is given by the vector space of vertical maps in the following diagram:
\begin{equation}
\begin{tikzpicture}[scale=2, row sep=3em, column sep=2.8em]
\node (A) at (0,0) {$S^{\oplus 2}$};
\node (B) at (1.5,0) {$S^{\oplus 2}$};
\node (D) at (0, -1) {$S^{\oplus 2}$};
\node (C) at (-1.5, -1) {$S^{\oplus 2}$};

\draw (A) edge[->] node[auto, font=\scriptsize]{$T$} (B);
\draw (C) edge[->] node[auto, font=\scriptsize]{$T$} (D);
\draw (A) edge[->] node[auto, font=\scriptsize]{$\delta T$} (D);
\draw (A) edge[->, dashed] node[near end, above=.6em, font=\scriptsize]{$-\mathfrak{g}_{\overline{\rm D9}}$} (C);
\draw (B) edge[->, dashed] node[near start, below=.6em, font=\scriptsize]{$\mathfrak{g}_{\rm D9}$} (D);

\end{tikzpicture}
\end{equation}
Here we have used suggestive notation to recognize the physical meaning of the various elements in the diagram. The two rows are copies of the same two-term complex. The vertical map goes from the anti-D9-stack on top to the D9-stack below, and hence is `of the same species' as the tachyon $T$ itself. This lends intuition to the idea that $\delta T$ can be thought of as a deformation of $T$. The left and right homotopy maps are easily recognized as automorphisms of the anti-D9 and D9 stacks, respectively.

Now let us compute the spectrum in this language. First of all, any dependence on the coordinate $z$ in $\delta T$ can be easily eliminated via a suitable homotopy. Given 
\be 
\delta T = z \, M\,,
\ee
for some matrix $M$, we can always define 
\be
\mathfrak{g}_{\rm D9} = -\mathfrak{g}_{\overline{\rm D9}} = \tfrac{1}{2} M\,,
\ee
such that 
\be 
\delta T = \mathfrak{g}_{\rm D9} \, T- T \, \mathfrak{g}_{\overline{\rm D9}} \sim 0\,.
\ee
This just tells us that the spectrum of fluctuations of the tachyon is localized on the brane at $z=0$. Hence, we can identify $\delta T$ with the Higgs field $\Phi$ mentioned in the introduction. Now we can study $z$-independent fluctuations modulo homotopies with $\mathfrak{g}_{\overline{\rm D9}} = +\mathfrak{g}_{\rm D9}$, which reduces to looking for adjoint matter in the $U(2)$ theory.

Given two branes represented by complexes $A_\bullet$ and $B_\bullet$, we can compute the chiral and anti-chiral bifundamental spectrum as Ext$^1(A_\bullet, B_\bullet)$ and Ext$^1(B_\bullet, A_\bullet)$, respectively.

Finally, let us introduce one more concept: The \emph{cone construction}. Just as we can define objects as kernels or cokernels of maps between sheaves, so can we use morphisms between complexes of sheaves to define other objects. This will allow us to bind a bigger variety of branes together, as we will see later on. Given two complexes $(A_\bullet, B_\bullet)$, and a morphism $m_\bullet$ between them
\begin{equation}
\begin{tikzpicture}[baseline=(current bounding box.center), row sep=3em, column sep=2.8em, text height=1.5ex, text depth=0.25ex]
\node (upleft) at (-1.5,0) {$\cdots$};
\node (downleft) at (-1.5, -1.4) {$\cdots$};
\node (upright) at (7.5,0) {$\cdots$};
\node (downright) at (7.5,-1.4) {$\cdots$};
\node (A) at (0,0) {$A_{i-1}$};
\node (B) at (3,0) {$A_{i}$};
\node (C) at (6,0) {$A_{i+1}$};
\node (D) at (0,-1.4) {$B_{i-1}$};
\node (E) at (3,-1.4) {$B_i$};
\node (F) at (6,-1.4) {$B_{i+1}$};
\path[->,font=\scriptsize]
(A) edge node[auto] {$d^A_{i}$} (B)
(B) edge node[auto] {$d^A_{i+1}$} (C)
(D) edge node[auto] {$d^B_{i-1}$} (E)
(E) edge node[auto] {$d^B_{i}$} (F)
(A) edge node[auto] {$m_{i-1}$} (D)
(B) edge node[auto] {$m_i$} (E)
(C) edge node[auto] {$m_{i+1}$} (F)
(upleft) edge (A)
(downleft) edge (D)
(C) edge (upright)
(F) edge (downright)
;
\end{tikzpicture}
\end{equation}
we define a third complex $C_\bullet$, called the \emph{mapping cone} of $m_\bullet$ as the following complex
\begin{equation}
\begin{tikzpicture}[baseline=(current bounding box.center), row sep=3em, column sep=2.8em, text height=1.5ex, text depth=0.25ex]
\node (upleft) at (-1.5,0) {$\cdots$};
\node (downleft) at (-1.5, -1.4) {$\cdots$};
\node (upright) at (7.5,0) {$\cdots$};
\node (downright) at (7.5,-1.4) {$\cdots$};
\node (A) at (0,0) {$A_{i}$};
\node (B) at (3,0) {$A_{i+1}$};
\node (C) at (6,0) {$A_{i+2}$};
\node (p1) at (0,-.7) {$\oplus$};
\node (p2) at (3,-.7) {$\oplus$};
\node (p3) at (6,-.7) {$\oplus$};
\node (D) at (0,-1.4) {$B_{i-1}$};
\node (E) at (3,-1.4) {$B_i$};
\node (F) at (6,-1.4) {$B_{i+1}$};
\path[->,font=\scriptsize]
(A) edge node[auto] {$-d^A_{i}$} (B)
(B) edge node[auto] {$-d^A_{i+1}$} (C)
(D) edge node[auto] {$d^B_{i-1}$} (E)
(E) edge node[auto] {$d^B_{i}$} (F)
(A) edge node[above] {$m_i$} (E)
(B) edge node[above] {$m_{i+1}$} (F)
(upleft) edge (A)
(downleft) edge (D)
(C) edge (upright)
(F) edge (downright)
;
\end{tikzpicture}
\end{equation}
By introducing more notions, one can make sense of the notion that $C_\bullet$ is the cokernel of $A_\bullet \stackrel{m_\bullet}\longrightarrow B_\bullet$, i.e. $C_\bullet \equiv B_\bullet/m_\bullet(A_\bullet)$.
This construction allows one to create a brane as the difference between two other branes.

\section{T-branes as coherent sheaves} \label{sec:coherentsheaves}
In this section, we will introduce coincident as well as intersecting D7-branes that form bound states via the condensation of open strings with unequal Chan Paton labels at their extremities. For coincident branes, such setups were introduced in \cite{Gomez:2000zm, Donagi:2003hh}. Later on, in \cite{Heckman:2010qv} and \cite{Donagi:2011jy}, more general bound states were discovered that included intersecting branes, and in \cite{Cecotti:2010bp} a systematic analysis of such systems was carried out. The term ``T-branes'' was coined to describe 7-brane systems where the worldvolume adjoint Higgs field is given a vev that cannot be entirely captured by its characteristic polynomial.

The purpose of this chapter is to recast the treatment of T-branes into the language of coherent sheaves. We will present a few basic examples exhibiting unexpected behavior. We will first use the standard language, in terms of a background Higgs field on the worldvolume of the D7-branes. We will then rewrite everything in terms of tachyon condensation of D9/anti-D9-brane pairs. This point of view not only drastically simplifies the information, but also clarifies the apparently unusual behavior of systems described as T-branes.

\subsection{Gauge group breaking via nilpotent Higgsing} \label{sec:basictbrane}

\subsubsection*{Higgs picture}

The following is the simplest example of a T-brane \cite{Cecotti:2010bp}. Take two coincident, flat D7-branes: The theory has a $U(2)$ gauge symmetry with a complex adjoint scalar $\Phi$, which parametrizes the two transverse directions to the brane.
Denoting by $\sigma_i$, for $i=1,2$ or $3$, the standard Pauli matrices, switching on a vev along any $\sigma_i$ corresponds to separating the two branes in this transverse space. One way to see this is to realize that the algebraic equation describing the brane system is given by the characteristic equation of the adjoint scalar. Denoting by `$z$' the complex transverse coordinate, the equation is then given by:
\begin{equation}
\det(\mathbb{1} z-\Phi)= z^2+\det(\Phi) = 0\,,
\end{equation}
whereby we used the tracelessness of $\Phi$. The point is that any vev of $\Phi$ that admits a diagonalization, i.e. $\Phi = \sigma_i$, will amount to separating the branes, and breaking the gauge group down to $U(1) \times U(1)$. However, since $\Phi$ is complex, we can switch on a vev along a nilpotent direction, such as
\begin{equation} \label{nilpotenthiggs}
\Phi = \begin{pmatrix} 0 & 1 \\ 0 & 0 \end{pmatrix}\,.
\end{equation}
The characteristic equation is unaffected by this, but the gauge group that leaves this vev invariant is the overall $U(1)$. The stack of branes behaves as a bound state, with only a `center of mass' $U(1)$. Note, that this system is not onshell as an eight-dimensional background as it violates the equation of motion $[ \Phi, \Phi^\dagger]=0$. Nevertheless, the example exhibits properties that are universal, hence its interest.

One might wonder how this data is encoded in an F-theory, or dual M-theory picture. This particular vev corresponds to a coherent state of strings stretching from one brane to the other, and hence we expect it to lift to a coherent state of light M2-branes in the dual M-theory picture, as opposed to some geometric modulus of the Calabi-Yau (CY). Indeed, we can easily construct the relevant patch of the elliptically fibered CY manifold for this setup. Since both branes are mutually local, there are only T-monodromies, no S-dualities around. Hence, only one 1-cycle of the elliptic fiber will collapse, and the other 1-cycle will never mix with it. Hence, we can focus on a patch of the torus in the shape of a cylinder. More precisely, consider the following geometry:
\begin{equation}
u v = \det(\mathbb{1} z-\Phi) \, \qquad {\rm in}\qquad \mathbb{C}^3 \times \mathbb{R}^{1,7}\,,
\end{equation}
with the $\mathbb{C}^3$ parametrized by $u, v, z$. For $\Phi=0$, we have $u v = z^2$.
Over each point away from $z=0$, we have a $\mathbb{C}^*$-fiber, and over $z=0$, the fiber degenerates to the union of two lines. The whole fibration is in the shape of an A$_1$ surface singularity. The key point here is that, if we would turn on a vev for $\Phi$ along the Cartan of $SU(2) \subset U(2)$, the surface singularity would get smoothed by deformation.
\be
uv = z^2 \qquad \longrightarrow \qquad uv = z^2+\epsilon^2\,.
\ee
for $\Phi = {\rm diag}(\epsilon, -\epsilon)$. On the other hand, for the nilpotent choice \eqref{nilpotenthiggs}, the equation remains unaffected. Hence, this Higgs branch does not correspond to a geometric modulus of the F/M-theory compactification. Rather, it corresponds to a coherent state of M2-branes wrapped on the vanishing sphere of the A$_1$ singularity. This point of view will be further developed in our companion paper \cite{Collinucci:2014taa}.

\subsubsection*{Tachyon condensation picture}

We are now ready to study our first T-brane using the tachyon condensation picture. Set the background tachyon to the value
\be T = \bp z & 1 \\ 0 & z \ep\,. \ee
This is equivalent to taking an $U(2)$ stack of D7-branes, and turning on a nilpotent Higgs as in \eqref{nilpotenthiggs}. However, now our efforts will pay off, and we will see that the tachyon condensation picture makes the problem trivial. By performing a finite gauge transformation on the stack of anti-D9's and another one on the stack of D9's, we can bring our tachyon to a diagonal form:
\be\label{D9-D9gaugeu2}
T \longrightarrow  \tilde T= \bp z & -1 \\ 1 & 0 \ep T \bp 1 & 0 \\ -z & 1 \ep = \bp z^2 & 0 \\ 0 & 1 \ep\,.
\ee
Note that these transformations have constant non-zero determinant, which is why they are automorphisms. The crucial reason why this maneuver is possible is that the off-diagonal entry of the tachyon is constant. From the $U(2)$ Higgs point of view, $\Phi$ cannot be diagonalized. There, the only available gauge transformations are in the adjoint of $U(2)$. From the tachyon condensation point of view, the crucial advantage is that there are two independent gauge transformations one can perform on the left and right, and these need not be each other's inverses.

Now let us interpret our new tachyon $\tilde T$: It is reducible, which means it represents two decoupled D7-branes. However, the $(2,2)$ entry, being invertible, represents an empty brane. In other words, the complex
\bcd
S \rar{\cdot 1} & S \,.
\ecd
has no cokernel. Physically, a constant non-zero tachyon implies total D9/anti-D9 annihilation, leaving no D7 behind. Hence, this whole system is simply equivalent to
\bcd
S \rar{z^2} &S \,. \ecd
We can simply say that $T \cong \tilde T \cong z^2$. This is a bound state of two coincident branes. It is now manifest that the system just has an overall decoupled $U(1)$ gauge symmetry inherited from the D9 and anti-D9 gauge symmetries:

\begin{equation} 
\begin{tikzpicture}[scale=2]
\node (a) at (0,0) {$S$};
\node[right=of a] (b) {$S$};

\draw (a) edge[->] node[auto, font=\scriptsize]{$z^2$} (b);
\draw (a) edge[->, loop below] node[font=\scriptsize]{$g_{U(1)}$}  (a);
\draw (b) edge[->, loop below] node[font=\scriptsize]{$\tilde g_{U(1)}$}  (b);

\end{tikzpicture}
\end{equation}
with $g_{U(1)} = \tilde g_{U(1)}$.

\subsection{Phantom curve}\label{PhantomCurveSection}

\subsubsection*{Higgs picture}

Another slightly more interesting example is that of two intersecting branes. Two intersecting branes can be written as a `Higgsed' $U(2)$ theory. Take two coincident D7-branes at $z=0$, and switch on a position dependent vev along, say, $\sigma_3$:
\begin{equation}
\Phi_0 = x \, \sigma_3\,,
\end{equation}
where $x$ is a longitudinal worldvolume coordinate on the D7-stack.
This breaks the $SU(2)\subset U(2)$ down to a `relative' $U(1) \subset SU(2)$. This relative group is generated by the difference of the generators of the $U(1)$'s of the two intersecting branes. 

Nevertheless, at the `matter' curve $z=x=0$, the gauge group enhances back to $U(2)$. At this locus, there is trapped matter that can be analyzed by decomposing the adjoint of $U(2)$ in terms of the Cartan `relative' $U(1)$. The interesting states are the off-diagonal fluctuations of this Higgs, which are charged $\pm 2$ w.r.t. to it. These are the bifundamental strings stretching from one brane to the other. The characteristic equation is simply $(z+x)\,(z-x)$, which clearly shows the reducible geometry of the total system. 

It is interesting to keep track of the F/M-theory lift of this system. A particular patch of the corresponding elliptic fibration is the following
\be
u v =(z+x)\,(z-x)\,.
\ee
This is a family of conifold singularities fibered over the `matter curve' at $z=x=0$. By performing a small resolution, we detect a vanishing $\P^1$. An M2 wrapping this sphere corresponds to the M-theory lift of a bifundamental string of charge $+2$, and an anti-M2 corresponds to an oppositely oriented string of charge $-2$.
Note, that by insisting on a smooth F-theory manifold, we are forced to choose one of the two small resolutions in a very unnatural way. The two choices are related by a flop transition. From the 3d gauge theory point of view, this transition corresponds to a Weyl reflection on the enhanced $U(2)$ which exchanges the two Coulomb branches of the theory \cite{Diaconescu:1998ua, Grimm:2011fx, Cvetic:2012xn, Hayashi:2013lra}. In our companion paper \cite{Collinucci:2014taa}, we will propose a language for describing the F-theory uplifts of intersecting branes in the singular phase. This will facilitate the exploration of T-brane backgrounds, which would otherwise be prohibited once a desingularization has taken place.

Now, let us turn on a more interesting background on this intersecting brane system:
\be \label{phantomhiggs}
\Phi = \bp x &  1 \\ 0 & -x \ep \,.
\ee
The characteristic equation and therefore the F-theory uplift of this background is exactly the same as before. However, the gauge group is now completely broken. By switching on the off-diagonal term, we have condensed a bifundamental string, and now the two D7-branes are bound together and behave as one brane. There are no charged states because there is only a decoupled `center of mass' $U(1)$. Hence, all degrees of freedom are center of mass displacements of this brane. The fact that the geometry of the total system falsely displays a matter curve at $z=x=0$, even though no matter is localized there, was dubbed `phantom curve' in 
\cite{Cecotti:2010bp}.

\subsubsection*{Tachyon condensation picture}

Describing the system with phantom curve in the tachyon condensation picture is remarkably simple. The tachyon corresponding to \eqref{phantomhiggs} is simply:
\be \begin{pmatrix} z+x & 1\\ 0 & z-x \end{pmatrix} \,. \ee
We will now easily see that this is a bound state, with just a `center of mass' $U(1)$, and therefore no charged localized matter. We simply perform independent gauge transformations on the D9's and anti-D9's as follows:
\be
T \longrightarrow  \tilde T= \bp z-x & \quad -1 \\ 1 &\quad  0 \ep T \begin{pmatrix} 1 & \quad 0 \\ -z-x & \quad 1 \end{pmatrix} = \begin{pmatrix} z^2-x^2 & 0 \\ 0 & 1 \end{pmatrix}\,.
\ee
Just as in the previous case, the `$1$' in the $(2,2)$ entry corresponds to an empty brane, so all we are left with is a bound state of two D7-branes at $z^2-x^2=0$ behaving as a single brane. The string that has condensed in order to bind these two branes changed the nature of the coherent sheaf, but did not displace its support. Such degrees of freedom were studied extensively in \cite{Donagi:2011jy}.

The standard tools to describe F-theory cannot handle this type of situation. See \cite{Anderson:2013rka}, however, for an alternate proposal. If one can only specify a geometry, and a $C_3$-form, one will not see this `gluing data'. One must pass to the M-theory picture, and include a coherent state of charged M2-branes wrapped on a vanishing cycle.
The main objective of our follow-up paper \cite{Collinucci:2014taa} is to provide a framework in which this extra information can be naturally and practically integrated into the F-theory picture.

\subsection{Nilpotent matter}
In this section we will study a case where the characteristic polynomial of the Higgs field completely misses massless modes. We will see that there are localized modes that can be explained by analyzing the D7-branes as coherent sheaves, and by discovering that they are in effect bound to an anti-D5-brane. The localized modes are remnants of the $\overline{{\rm D}5}/\overline{{\rm D}5}$ strings.

\subsubsection*{Higgs picture}

Consider the following Higgs field
\begin{equation} \label{nilpotentmatterhiggs}
\Phi = \begin{pmatrix} 0 & x \\ 0 & 0 \end{pmatrix}\begin{array}{c} \\ \end{array} \,.
\end{equation}
Let us first perform the standard analysis of the matter spectrum by parametrizing all possible Higgs fluctuations
\be
\delta \Phi = \bp \phi_0 & \phi_+ \\ \phi_- & - \phi_0 \ep
\ee
modulo linearized gauge transformations:
\be
\delta \Phi \sim \delta \Phi + [ \Phi, \chi ]
\ee
for some $2 \times 2$ matrix $\chi$
\be 
\chi  = \bp \chi_0 & \chi_+ \\ \chi_- & -\chi_0 \ep \begin{array}{c} \\ \end{array} \,.
\ee
The most general commutator gives
\be \begin{pmatrix} x\, \chi_- &  - 2 x\, \chi_0 \\ 0& -x\, \chi_- \end{pmatrix}\begin{array}{c} \\ \end{array} \,.
\ee
From this we deduce two modes localized on the curve $\{x=0\}$. In other words, if $(x, y)$ are longitudinal coordinates on the stack at $z=0$, then
\be 
\phi_+,\phi_0 \in \C[x, y]/(x)\,.
\ee
The system looks geometrically like a single stack of two D7-branes wrapping a surface, yet there appears matter localized on a curve.

\subsubsection*{Tachyon condensation picture}

Let us now use the tachyon condensation picture to demystify this peculiar behavior. We will find that the localized modes are due to an anti-D5-brane wrapping the locus $\{x=0\}$ inside the stack. The tachyon profile for the background in question looks like
\be
T = \begin{pmatrix} z & x \\ 0 & z \end{pmatrix} \,.
\ee
Defining the coordinate ring of the internal space $S = \C[x, y, z]$, we remind the reader that the 7-brane configuration is given by the \emph{cokernel sheaf} in the sequence:
\bcd 0 \rar & S^{\oplus 2} \rar{T} & S^{\oplus 2} \rar & {coker}(T) \rar & 0 \,.
\ecd
To understand the sheaf $coker(T)$, it suffices to study how the rank of the tachyon matrix $T$ varies along the internal space. For instance, at a generic point in the `internal' $\C^3$, $T$ has rank two. Hence, the cokernel of $T$ as a linear map is zero. In physical terms, after tachyon condensation, there is no D-brane at this locus.
At a slightly less generic locus, where $z = 0$, but $x \neq 0$, $T$ will have rank one. Hence, its cokernel jumps up to rank one. This means that there is a D-brane at this locus. Since the codimension is one, we interpret this as a D7-brane. In this case, it is a bound state of two coincident D7-branes as in \ref{sec:basictbrane}

Finally, at the even less generic locus where $z=0$ and $x=0$, the tachyon is zero, hence the cokernel has rank two. At this locus, we deduce that there is an object sitting on top of the D7-brane, raising the rank from one to two. Given that this is happening on an internal curve, this object must be either a D5 or an anti-D5. We will prove that is is actually an anti-D5-brane.
The situation is summarized as follows:

\begin{center}
\begin{tabular}{ccc}
\text{ideal} & \text{rank}(coker(T)) & \text{D-brane} \\
\hline
$(0)$ & $0$ & \text{none} \\
$(z)$ & $1$ & \text{D}7 \\
$(x,z)$ & $2$ & $\text{D}7+\overline{{\rm D}5}$
\end{tabular}
\end{center}

This heuristic picture based on analyzing the rank of the tachyon matrix tells us roughly what to expect. But in order to get a more precise picture, we must now describe this system in terms of complexes. We will rewrite the complex describing our system
\bcd[ampersand replacement=\&, column sep=large] 
S^{\oplus 2} \rar{\bp z & x \\ 0 & z \ep} \& S^{\oplus 2}
\ecd
as a bound state of a pure D7 system and an anti-D5. The D7 will be a bound state given by
\bcd[ampersand replacement=\&, column sep=large] 
S^{\oplus 2} \rar{\bp z & 1 \\ 0 & z \ep} \& \underline{S^{\oplus 2}}
\ecd
which has no localized matter whatsoever. The complex for an anti-D5 is given by
\bcd[ampersand replacement=\&, column sep=large]
S \rar{\bp -x \\ z \ep} \& \underline{S^{\oplus 2}} \rar{\bp z & x \ep} \& S \,.
\ecd
Note, that in a complex, the zeroth position is a matter of convention. In this case, we have underlined the objects in position zero. This choice fixes which objects are to be thought of as D9's, and which as anti-D9's. Our particular choice is such that the first complex corresponds to a D7-brane, and the second complex to an anti-D5-brane. 

Consider the following morphism in Ext$^1(D7, \overline{D5})$ between these two objects
\begin{equation}
\begin{tikzpicture}[scale=2]
\node (D71) at (1,1) {$S^{\oplus 2}$};
\node (D72) at (2.5,1) {$\underline{S^{\oplus 2}}$};
\node (D51) at (0,0) {$S$};
\node (D52) at (1,0) {$\underline{S^{\oplus 2}}$};
\node (D53) at (2.5,0) {$S$};
\node (D7) at (4,1) {D$7$};
\node (D5bar) at (4,0) {$\overline{D5}$};

\path[->,font=\scriptsize]
(D71) edge node[auto] {$\bp -z & -1 \\ 0 & -z \ep$} (D72)
(D51) edge node[below] {$\bp -x \\ z \ep$} (D52)
(D52) edge node[below] {$\bp z & x \ep$} (D53)
(D71) edge[thick, red]  node[black, auto] {$\bp 0 & 1 \\ 0 & 0 \ep$} (D52)
(D72) edge[thick, red]  node[black, auto] {$\bp 0 & 1 \ep$} (D53)
(D7) edge[thick, red] (D5bar)
;
\end{tikzpicture} 
\end{equation}
The upper complex represents a bound state of two D7-branes. The lower complex is an anti-D5 shifted by one to the left.
Note that this morphism is constant along the worldvolume of the D7-brane, so it does not generate any vortices. Now, in order to form a bound state, we will take a \emph{mapping cone} w.r.t. the morphism. Remember that the mapping cone w.r.t. a morphism between two complexes $A_\bullet \rightarrow B_\bullet$ means creating an object that is roughly a difference between these two objects. In this case, the cone has the following form

\begin{equation}
\begin{tikzpicture}[scale=3]
\node (D71) at (0,1) {$S^{\oplus 2}$};
\node (D72) at (1,1) {$\underline{S^{\oplus 2}}$};
\node (D51) at (0,0) {$S$};
\node (D52) at (1,0) {$\underline{S^{\oplus 2}}$};
\node (D53) at (2,0) {$S$};
\node (p1) at (0, .5) {$\oplus$};
\node (p2) at (1, .5) {$\oplus$};

\path[->,font=\scriptsize]
(D71) edge node[auto] {$\bp z & 1 \\ 0 & z \ep$} (D72)
(D51) edge node[below] {$\bp -x \\ z \ep$} (D52)
(D52) edge node[below] {$\bp z & x \ep$} (D53)
(D71) edge[bend right] node[auto] {$\bp 0 & 1 \\ 0 & 0 \ep$} (D52)
(D72) edge[bend right] node[auto] {$\bp 0 & 1 \ep$} (D53)
;
\end{tikzpicture} 
\end{equation}
This whole system can be written as the following complex

\begin{equation}
\begin{tikzpicture}[scale=3]
\node (A) at (0,0) {$S^{\oplus 3}$};
\node (B) at (1,0) {$\underline{S^{\oplus 4}}$};
\node (C) at (2,0) {$S$};

\path[->,font=\scriptsize]
(A) edge node[auto] {$\bp z & 1 & 0 \\0 &  z & 0  \\ 0 & 1 & -x \\ 0 & 0 & z \ep$} (B)
(B) edge node[auto] {$\bp 0 & 1 & z & x \ep$} (C)
;
\end{tikzpicture}
\end{equation}
By performing appropriate automorphisms on the various terms of the sequence and eliminating parts with trivial cohomology, i.e. things of the form $S \stackrel{\cdot 1}\longrightarrow S$, we arrive at the form

\bcd[ampersand replacement=\&, column sep=large]
S^{\oplus 2} \rar{\bp z & x \\ 0 & z \ep}\&  \underline{S^{\oplus 2}}\,,
\ecd
which is the system we wanted. Now the origin of the strange matter curve at $z=x=0$ is elucidated: There is no second D7-brane intersecting our system on the curve; on the other hand, there was an anti-D5 situated precisely at that curve. The localized nilpotent matter are the remnants of the $\overline{{\rm D}5}/\overline{{\rm D}5}$ strings that are still localized after the $\overline{{\rm D}5}$ has formed a bound state with the D7.

Note, that if we take D-terms into account, the anti-D5 will dissolve over the worldvolume of the D7, and the wave function of this matter will spread out. This can only be seen by passing to unitary frame.

\subsection{Point-like matter} \label{sec:pointlike}
Finally, let us introduce another peculiar phenomenon that can occur when non-Abelian degrees of freedom are turned on. The standard picture in IIB string theory has a clearly established hierarchy of gauge theory structures arranged according to codimension: Gauge groups reside on divisors, matter on intersection curves, and Yukawa couplings on triple intersections of 7-branes.
In \cite{Cecotti:2010bp}, a case was found where the matter seems to be localized on a point as opposed to a curve, which defies usual intuition. In this section, we will treat a simpler setup that captures the same behavior, and through the tachyon condensation picture, we will demystify the origin of the point-like matter.

\subsubsection*{Higgs picture}
Let us start again with an $U(2)$ stack of D7-branes at $z=0$ with longitudinal complex coordinates $(x, y)$, and switch on the following vev for the adjoint Higgs:
\be \label{higgspointlike}
\Phi = \bp x & y \\ 0 & -x \ep \,.
\ee
This Higgs cannot be diagonalized. It breaks the $U(2)$ down to the overall $U(1)$ everywhere outside the locus $y=0$. On the curve $y=0$, it enhances to a $U(1) \times U(1)$. At the intersection of the curves $x=0$ and $y=0$, it enhances all the way back to $U(2)$. This tells us that there might be trapped matter at the various enhancement loci. The way to analyze this is by parametrizing all possible fluctuations of the Higgs field 
\be
\delta \Phi = \bp \phi_0 & \phi_+ \\ \phi_- & - \phi_0 \ep 
\ee
modulo linearized gauge transformations:
\be
\delta \Phi \sim \delta \Phi + [ \Phi, \chi ]
\ee
for some $2 \times 2$ matrix $\chi$
\be 
\chi  = \bp \chi_0 & \chi_+ \\ \chi_- & -\chi_0 \ep \,.
\ee
The most general commutator gives
\be \begin{pmatrix} y\, \chi_- & \quad 2 x\, \chi_+-2 y\, \chi_0 \\ 2 x\, \chi_-& -y\, \chi_- \end{pmatrix} \,.
\ee
Choosing only $\chi_+$ and $\chi_0$ non-zero, we discover that the fluctuation $\phi_+$ is entirely localized at the point $x=y=0$. More precisely, $\phi_+$ is a polynomial in $x$ and $y$ modulo the ideal $(x,y)$
\be \phi_+ \in \C[x,y]/(x,y) \,. \ee 
This is a simple instance of \emph{point-like matter}. The system looks geometrically like a pair of intersecting branes, yet there is a field localized at a point, defying common expectation.

\subsubsection*{Tachyon condensation picture}
In this section, we will use the tachyon condensation picture to understand why the background \eqref{higgspointlike} exhibits such peculiar behavior. We will discover that there is actually an anti-D3-brane lurking, giving rise to modes localized at a point.

The tachyon profile corresponding to \eqref{higgspointlike} is
\be
T = \begin{pmatrix} z+x & y \\ 0 & z-x \end{pmatrix} \,.
\ee
Defining the coordinate ring of the internal space $S = \C[x,y,z]$, we remind the reader that the 7-brane configuration is given by the \emph{cokernel sheaf} in the sequence:
\bcd 0 \rar & S^{\oplus 2} \rar{T} & S^{\oplus 2} \rar & {coker}(T) \rar & 0 \,.
\ecd
To get a heuristic picture of what the sheaf $coker(T)$ is made of, it suffices to study how the rank of the tachyon matrix $T$ varies in the internal space. For instance, at a generic point in $\C^3$, $T$ has rank two. This means that the cokernel of $T$ as a linear map is zero. The physical interpretation is that, after tachyon condensation, there is no D-brane at this locus.
At a slightly less generic locus, where $(z+x)\, (z-x) = 0$, but $y \neq 0$, $T$ will have rank one. Hence, its cokernel jumps up to rank one. This means that there is a D-brane at this locus. Since the codimension is one, we interpret this as a (reducible) D7-brane.

Finally, at the locus $z=x=y=0$, $T$ has rank zero, bumping up the cokernel to rank two. Therefore, there is point-like brane on top of the D7-brane at this point, which we will show is an anti-D3-brane. We summarize this below:

\begin{center}
\begin{tabular}{ccc}
\text{ideal} & \text{rank}(coker(T)) & \text{D-brane} \\
\hline
$(0)$ & $0$ & \text{none} \\
$(z+x)$ & $1$ & \text{D}$7_+$ \\
$(z-x)$ & $1$ & \text{D}$7_-$ \\
$(x,y,z)$ & $2$ & $\text{D}7+\overline{{\rm D}3}$
\end{tabular}
\end{center}
This clarifies the origin of the point-like matter: It corresponds to remnants of the $\overline{{\rm D}3}/\overline{{\rm D}3}$ strings.

Let us show this embedded anti-D3-brane more explicitly by actually creating a bound state between the two D7's at $z+x=0$ and $z-x=0$, and the anti-D3 at $x=y=z=0$. Since we are working on an affine space, the homotopy category is equivalent to the derived category of coherent sheaves, so we only need to worry about homotopies.
A string from the D$7_+$ at $z+x=0$ to the anti-D3 binds these two together, while a string from the anti-D3 to the other D$7_-$ at $z-x=0$ binds these two together. More precisely, consider the following two morphisms in Ext$^1(D7_+, \overline{D3})$ and Ext$^1(\overline{D3}, D7_-)$, respectively:

\begin{equation}\label{AntiD3bound}
\begin{tikzpicture}[scale=1.8]
\node (D711) at (4,2) {$S$};
\node (D712) at (6,2) {$\underline{S}$};
\node (D7+) at (8,2) { \, D$7_+$};
\node (D31) at (1,1) {$S$};
\node (D32) at (2.5,1) {$S^{\oplus 3}$};
\node (D33) at (4,1) {$\underline{S^{\oplus 3}}$};
\node (D34) at (6,1) {$S$};
\node (D3bar) at (8,1) {$\overline{D3}$};
\node (D721) at (1,0) {$S$};
\node (D722) at (2.5,0) {$\underline{S}$};
\node (D7-) at (8,0) { \, D$7_-$};

\path[->,font=\scriptsize]
(D711) edge node[auto] {$(z+x)$} (D712)
(D31) edge node[auto] {$-X_2$} (D32)
(D32) edge node[auto] {$-X_1$} (D33)
(D33) edge node[auto] {$-X_0$} (D34)
(D721) edge node[below] {$(z-x)$} (D722)
(D712) edge[thick, blue] node[black, auto] {$ 1 $}(D34)
(D711) edge[thick, blue] node[black, auto] {$-\bp 1 \\ 0 \\ 1 \ep$}(D33)
(D31) edge[thick, red] node[black, auto] {$1$}(D721)
(D32) edge[thick, red] node[black, auto] {$\bp -1 & 0 & 1 \ep$}(D722)
(D7+) edge[blue, thick] (D3bar)
(D3bar) edge[red, thick] (D7-)
;
\end{tikzpicture}
\end{equation}
where
\be \nonumber
X_0 \equiv \bp x & y & z \ep\,, \quad X_1 \equiv \bp -y & -z & 0 \\ x & 0 & -z \\ 0 & x & y \ep\,, \quad X_2 \equiv \bp z \\ -y \\ x \ep\,.
\ee
The top complex represents the D7$_+$, the middle complex is an anti-D3 shifted by one to the left, and the bottom complex is the D7$_-$ shifted two places to the left. The rightmost column schematically summarizes the idea.

Now, in order to create a bound state which corresponds to `condensing' the blue D7$_+$/$\overline{D3}$ strings, and the red $\overline{D3}$/D7$_-$ strings, we take the mapping cones with respect to both morphisms. The result is the following:
\begin{equation}
\begin{tikzpicture}[scale=2]
\node (D711) at (2,2) {$S$};
\node (D712) at (4,2) {$\underline{S}$};
\node (p1) at (2,1.5) {$\oplus$};
\node (p2) at (4,1.5) {$\oplus$};
\node (D31) at (0,1) {$S$};
\node (D32) at (2,1) {$S^{\oplus 3}$};
\node (D33) at (4,1) {$\underline{S^{\oplus 3}}$};
\node (D34) at (6,1) {$S$};
\node (p3) at (2,.5) {$\oplus$};
\node (p4) at (4,.5) {$\oplus$};
\node (D721) at (2,0) {$S$};
\node (D722) at (4,0) {$\underline{S}$};

\path[->,font=\scriptsize]
(D711) edge node[auto] {$(z+x)$} (D712)
(D31) edge node[below] {$X_2$} (D32)
(D32) edge node[below] {$X_1$} (D33)
(D33) edge node[below] {$X_0$} (D34)
(D721) edge node[below] {$(z-x)$} (D722)
(D712) edge[bend right] node[auto, near start] {$ -1 $}(D34)
(D711) edge[bend right] node[auto] {$\bp 1 \\ 0 \\ 1 \ep$}(D33)
(D31) edge[bend right] node[auto] {$1$}(D721)
(D32) edge[bend right] node[auto] {$\bp -1 & 0 & 1 \ep$}(D722)
;
\end{tikzpicture}
\end{equation}
We have chosen the zero position of the complex (displayed by the underlined objects), such that we have two D7-branes and one anti-D3-brane. After performing appropriate automorphisms on the various objects, we arrive at the desired complex:
\bcd[ampersand replacement=\&, column sep=huge]
S^{\oplus 2} \rar{\bpp z+x & y \\ 0 & z-x \epp} \& \underline{S^{\oplus 2}}\,.
\ecd
This completely elucidates the origin of the mysterious point-like matter. There is an embedded anti-D3 bound to the D7-stack. The point-like matter corresponds to D7/D7 strings, whose origin are the $\overline{{\rm D}3}/\overline{{\rm D}3}$-strings, before the $\overline{{\rm D}3}$ was bound to the D7.  It is natural to expect that, whenever T-branes display such peculiar behavior, there will be an explanation in terms of embedded lower D-branes present in the system.

\section{Spectrum of T-brane backgrounds}\label{TspectrumSec}

In this section we discuss other peculiar features of the matter spectrum of T-brane backgrounds not related to lower-dimensional branes. Using the tachyon condensation picture, we will analyze fluctuations around various backgrounds, as reviewed in section \ref{TachyonCondensationReview}. This method will allow us to easily visualize the gauge symmetry breaking patterns induced by gluing data.

Let us work in affine space, with $S=\C[z]$ the ring of functions in one complex variable. In other words, we consider eight-dimensional theories. A stack of $n$ D7-branes carrying an unbroken $U(n)$ gauge symmetry is described by the following tachyon background
\be
\begin{tikzcd}
S^{\oplus n} \arrow{r}{z \1_n} & 
S^{\oplus n}\,.
\end{tikzcd}
\ee
Starting from this system, one can create several T-brane configurations, corresponding to turning on vevs for the various open strings with \emph{unequal} Chan-Paton factors. Because the branes of the stack are all indistinguishable, we can limit ourselves, without loss of generality, to tachyons of the following form
\be\label{generalTun}
T=\left(\begin{array}{ccccc} z&*& & &\\ & z & * & & 0 \\ & & \ddots&\ddots& \\ &0&&z&*\\&&&&z \end{array}\right)\,\begin{array}{c}\\ \\ \\ \\  \end{array} \,.
\ee 
In other words, we can focus on just the open strings stretching between two consecutive D7-branes. Moreover, any $z$-dependence in their vevs can be gauged away. This means that the $*$'s in \eqref{generalTun} are all constants, and we can take them to be either $0$ or $1$. We thus see that the possible \emph{Jordan block} structures of the $n\times n$ tachyon give us all inequivalent ways of binding together two or more D7-branes in the stack. Tachyons differing by permutations of Jordan blocks are gauge equivalent. Therefore we can classify the inequivalent T-brane backgrounds of a $U(n)$ system just by a set of integers, $\{m_i\}_{i=1,\ldots,n}$, where $m_i$ is defined to be the \emph{multiplicity} with which the $i\times i$ Jordan block appears in \eqref{generalTun}. Having no T-brane corresponds to $m_1=n$ and all others equal to zero. This classification is well-known in group theory under the name of `nilpotent orbits', and it extends to groups of the ${\bf D}$ and ${\bf E}$ types \cite{Collingwood}. According to the general theory, starting from $U(n)$, the gauge group left unbroken by a T-brane configuration corresponding to the set $\{m_i\}$ is $\Pi_iU(m_i)$. To simplify computations, every time a $i\times i$ Jordan block appears, we can make D9/anti-D9 independent gauge transformations of the type \eqref{D9-D9gaugeu2} and throw away empty summands, in order to reduce the whole block to the $1\times 1$ matrix $z^i$.

The aim is now to find the massless fluctuations about a given T-brane background. For concreteness, let us work in the $n=4$ case, which already shows all the interesting features of T-branes of $U(n)$ systems. There are four inequivalent T-brane backgrounds we can construct: 
\begin{subequations}
  \begin{align}
   \{2,1,0,0\} : &\quad T=\bpp z &1&0&0\\ 0&z&0&0 \\ 0&0&z&0 \\0&0&0&z   \epp \simeq  \bpp z^2&0&0\\ 0&z&0\\ 0&0&z \epp & \hspace{-1cm}U(1)\times U(2) \;, & \label{2100}  \\ 
     \{0,2,0,0\}  : & \quad T= \bpp z &1&0&0\\ 0&z&0&0 \\ 0&0&z&1 \\0&0&0&z   \epp \simeq  \bpp z^2&0\\ 0&z^2 \epp &\hspace{-1cm}U(2)  \;, & \label{0200}  \\
    \{1,0,1,0\}  : & \quad T= \bpp z &1&0&0\\ 0&z&1&0 \\ 0&0&z&0 \\0&0&0&z   \epp \simeq  \bpp z^3&0\\ 0&z \epp & \hspace{-1cm}U(1)^2 \;, & \label{1010} \\
  \{0,0,0,1\}  : & \quad T= \bpp z &1&0&0\\ 0&z&1&0 \\ 0&0&z&1 \\0&0&0&z   \epp \simeq  z^4 & \hspace{-1cm}U(1)   \;, & \label{0001}
  \end{align}
\end{subequations}
where the left-most column indicates the set of integers $\{m_i\}$ labeling each T-brane configuration and the right-most one shows the corresponding residual gauge group. For each of these configurations, we would now like to analyze the adjoint spectrum $\delta T$ by computing Ext$^1({\rm cok}T,{\rm cok}T)$. Let us discuss all cases in turn.

In the background \eqref{2100}, we have turned on a vev for the $1\to2$ string of the original system. The new system has less degrees of freedom: Intuitively, it has lost the fluctuations associated to all open strings whose decomposition in terms of $k\to k+1$ strings contain the $1\to2$ string, i.e. $1\to3$ (which can be written as $1\to2\to3$), $4\to2$ (which can be written as $4\to1\to2$), and also $1\to4$ and $3\to2$\footnote{They are easy to visualize in the tachyon matrix: They correspond to all entries which have a $1$ either in their row or in their column.}. It is interesting to zoom into the newly formed bound state between the first and the second D7-brane, and analyze its fluctuations. By going to the diagonal basis, we find two degrees of freedom in the tachyon variation $\delta T$
\be\label{deltaT1x1}
{\rm Ext}^1\left({\rm cok}(z^2),{\rm cok}(z^2)\right) \quad\Longrightarrow\quad \delta T= 2\alpha z-\beta \qquad \alpha,\beta\in\C\,.
\ee
Their distinct roles are actually more manifest in the old basis, with the $2\times2$ tachyon, where we find:
\be\label{deltaT2x2}
{\rm Ext}^1\left({\rm cok}\bpp z &1 \\ 0&z\epp\,,\,{\rm cok}\bpp z &1 \\ 0&z\epp\right) \quad\Longrightarrow\quad \delta T= \bpp\alpha &0 \\ \beta & \alpha\epp \qquad \alpha,\beta\in\C\,.
\ee
Using \eqref{D9-D9gaugeu2} and getting rid of the trivial summand, one can verify that \eqref{deltaT1x1} and \eqref{deltaT2x2} are indeed equivalent, modulo higher orders in the fluctuation.
In \eqref{deltaT2x2}, however, we clearly see that $\alpha$ is naturally associated to normal displacements of the bound state's center of mass, whereas $\beta$ shows up as a nilpotent fluctuation and corresponds to the former $2\to1$ string. This fluctuation may be given the interpretation of a `smoothing' degree of freedom of the support of the bound state.
Since the bound state enjoys only a $U(1)$ gauge symmetry, the fact that we find two complex scalars seems to be incompatible with minimal supersymmetry in eight dimensions, which requires just one complex scalar in an abelian vector multiplet. However, we have to remember that T-branes in eight dimensions are not onshell, as they are constant and violate the BPS equation $[\Phi, \Phi^\dagger]=0$.

Such constraint can instead be fulfilled in six or lower dimensions by including a suitable worldvolume flux. Therefore, we have a theory with at most eight supercharges, where vector multiplets do not have scalars. The fluctuations we are now discussing will fall in a number of adjoint-valued hypermultiplets (or chiral multiplets for the case of four supercharges), whose quantity will of course depend on the topology of the internal worldvolume of the bound state. The remaining fluctuations of the background \eqref{2100} correspond to the adjoint of the $U(2)$ system and a pair of bifundamental fields (i.e. strings stretching between the bound state and the $U(2)$ system).

The background \eqref{0200} is characterized by the independent formation of two bound states of the same nature, one between the first and the second brane and the other between the third and the fourth brane. Such a situation enjoys an enhancement of gauge symmetry from $U(1)\times U(1)$ to $U(2)$. The spectrum of fluctuations thus comprises the adjoint of $U(2)$, and in addition includes the smoothing degrees of freedom of the two bound states.

In the background \eqref{1010} a bound state of different nature is formed, namely one involving three branes, leaving apart a single brane of the original stack. Hence we have a system with two $U(1)$'s, and the spectrum includes their normal displacements together with a pair of bifundamental fields. These degrees of freedom are easily seen in the diagonal basis. Let us now zoom in on the $3$ by $3$ bound state. Here, we discover two additional fluctuations in the massless spectrum, which are associated to the former $3\to1$ string, i.e. the only string involving neither of the two turned on. Again these can be thought of as smoothing degrees of freedom, and are best identified in the basis with the non-diagonal tachyon:
\be\label{deltaT3x3}
{\rm Ext}^1\left({\rm cok}\bpp z &1&0 \\ 0&z&1\\0&0&z\epp\,,\,{\rm cok}\bpp z &1&0 \\ 0&z&1\\0&0&z\epp\right) \quad\Longrightarrow\quad \delta T= \bpp0 &0&0 \\ 0 & 0&0 \\ \beta z+\gamma &0&0\epp\qquad \beta,\gamma\in\C\,,
\ee
where we have omitted the center of mass (diagonal) fluctuation,  already accounted for.

Finally, in the background \eqref{0001}, all of the four branes of the original stack are bound together, thus breaking the gauge group all the way down to the $U(1)$ associated to their center of mass. With a computation similar to \eqref{deltaT2x2} and \eqref{deltaT3x3}, we find
\be\label{deltaT4x4}
\delta T= \bpp \alpha&&0 &&0&&0 \\ 0 && \alpha&&0&&0 \\ 0&&0&&\alpha&&0\\ \beta z^2+\gamma z + \delta &&0&&0&&\alpha\epp \qquad \alpha,\beta,\gamma,\delta\in\C\,,
\ee
where $\alpha$ represents the normal movements of the bound state, whereas $\beta,\gamma,\delta$ are its smoothing degrees of freedom.

Note that, for a bound state $z^k$, none of the $\delta T$'s we find yield a deformation proportional to $z^{k-1}$. The reason is that this type of fluctuation can be killed at first order by a change of coordinate. Indeed, suppose we have $z^k+\epsilon z^{k-1}$, with $\epsilon\ll 1$. Define $\tilde{z}={z+\epsilon/k}$. Then, in the new coordinate, we get $\tilde{z}^k+O(\epsilon^2)$, as claimed. The class of smoothings which our computation captures are non-trivial in this sense, and are known as `versal' deformations. Roughly speaking, they are defined as those hypersurface deformations which are not in the ideal of the gradient of the defining polynomial (the Jacobian ideal).

The same type of deformations shows up in the spectrum of fluctuations of the bound state of two intersecting D7-branes discussed in section \ref{PhantomCurveSection}. Indeed, for a bound state wrapping the locus $\{z^2-x^2=0\}$, with the usual Ext$^1$ computation in the $2\times2$ basis one finds
\be\label{deltaTxz}
T= \bpp z+x &1 \\ 0 & z-x\epp\quad\Longrightarrow\quad\delta T= \bpp \C[z-x] &0 \\ \epsilon & \C[z+x]\epp \qquad \epsilon\in\C\,,
\ee
where we see that the only smoothing degree of freedom appearing corresponds to the only possible versal deformation of the singular geometry, namely $z^2-x^2+\epsilon$.

\section{Stability in compactifications} \label{sec:stability}
So far in this paper, we have dealt only with topological B-branes, thereby ignoring any form of stability conditions. In doing this, we were able to effectively study the most peculiar features of T-branes, at the cost of ignoring conditions for the actual existence of such T-branes. Simply put, two D-branes will not bind together unless there is a tachyonic string stretched between them. Computing Ext groups allows one to determine the presence of stretched strings, but does not determine whether these are massive, massless or tachyonic.

\subsection{Review of $\Pi$-stability for D-particles}
We will briefly review the notion of $\Pi$-stability, which allows one to determine which B-branes can form bound states in the context of CY threefolds. The language we will employ is actually borrowed from a different setting, where one describes BPS D-particles in $\mathcal{N}=2, d=4$ compactifications of IIA on a CY threefold $X_3$. To access that setting, we simply get rid of the three spatial worldvolume dimensions of our branes. For instance, instead of thinking of two spacetime filling D7-branes that wrap four-cycles, one thinks of two D4-branes wrapping four-cycles, giving rise to two particles in four dimensions. 

IIA compactified on $X_3$ gives us a $\mathcal{N}=2, d=4$ effective theory. A Dp-brane wrapping an internal p-cycle can manifest itself as a 4d $1/2$-BPS particle that breaks the supersymmetry to an $\mathcal{N}=1_\xi \subset \mathcal{N}=2$. Here, $\xi$ parametrizes the relative phase of the linear combination of $\mathcal{N}=2$ generators that is preserved. 

To a given D-brane with RR-charge vector $\Gamma = (q_{D6}, q_{D4}, q_{D2}, q_{D0})$, we can attribute a `central charge' function $Z(\Gamma, t)$, where $t = B + i J$ is the complexified K\"ahler modulus. Two D-branes $\Gamma_1$ and $\Gamma_2$ will preserve the same $\mathcal{N}=1$ if the phases of their respective central charges are aligned \cite{Douglas:2000ah}:
\begin{equation}
\Gamma_1 \quad {\rm and}\quad \Gamma_2 \quad \text{mutually BPS} \qquad \Longleftrightarrow \qquad {\rm arg}\left(Z(\Gamma_1, t) \right) = {\rm arg}\left(Z(\Gamma_2, t) \right)\,.
\end{equation}
Note, that the property of being mutually BPS depends both on the charges \emph{and} the complexified K\"ahler modulus. The $\mathcal{N}=1 \subset \mathcal{N}=2$ preserved by a brane can be parametrized by the normalized argument\footnote{Although $\xi$ is a priori an angular variable, it turns out that we must `unfold' it, and let it run over $\mathbb{R}$ in order to get consistent results. See \cite{Aspinwall:2004jr} for an explanation. We will, however, not run into this issue in our applications.}
\begin{equation}
\xi \equiv \frac{1}{\pi}\, {\rm arg}(Z)\,.
\end{equation}

The central charge of a D-brane given by a coherent sheaf $\mathcal{S}$, can be computed at large volume with the following formula\footnote{We are now using the conventions in \cite{Douglas:2000ah}.}
\begin{equation} \label{centralcharge}
Z(\Gamma, t) = -\int_{X_3} e^{-t}\,{\rm ch}(\mathcal{S}) \sqrt{{\rm Td}(X_3)}\,,
\end{equation}
where all objects in the integrand are to be expanded as power series, and only the top-form components contribute to the integral. In this language, if we think of $\Gamma$ as a polyform
\begin{equation}
\Gamma = q_{D6} + {q_{D4}}^i \, D_i + {q_{D2}}_i \, C^i + q_{D0}\, \omega
\end{equation}
where, $D_i$ is a basis for $H^2(X_3, \mathbb{Z})$, $C^i$ is a basis for $H^4(X_3, \mathbb{Z})$, and $\omega$ is the volume form of $X_3$, then we can make the identification $\Gamma = {\rm ch}(\mathcal{S}) \sqrt{{\rm Td}(X_3)}$\,.

This formula for the central charge receives strong $\alpha'$ corrections from worldsheet instantons at generic points in moduli space that have to be computed by solving the Picard-Fuchs equation. However, we will mainly stay in the large volume regime to avoid this issue.

Let us now define the so-called $\Pi$-stability, in its specialization at large volume. Suppose we have a brane given by a coherent sheaf $\cS$ that fits in a short exact sequence
\begin{equation}
0 \rightarrow \cS_2 \longrightarrow \cS \longrightarrow \cS_1 \rightarrow 0\,.
\end{equation}
Then it is possible for $\cS$ to decay into $\cS_1$ and $\cS_2$ provided the following conditions are met:
\begin{enumerate}
\item We are near a locus in K\"ahler moduli space where the phases of $\cS_1$ and $\cS_2$ are aligned
\begin{equation}
\xi_1(t) = \xi_2(t)\,.
\end{equation}
Typically, this region will be a real codimension one locus in the moduli space. It is referred to as a \emph{wall of marginal stability}.

\item On the side of the wall where the open strings connecting $\cS_1$ and $\cS_2$ given by Ext$^1(\cS_1, \cS_2)$ are tachyonic, these will acquire a vev, thereby binding the two branes into a bound state $\cS$. This is the `stable side' of the wall. On the side where they are massive, the `unstable side', the bifundamental string has a vanishing vev, and the branes $\cS_1$ and $\cS_2$ are unbound. In other words, $\cS$ decays into these two branes, but the resulting vacuum breaks supersymmetry.
\end{enumerate}

The formula for the mass of an open bifundamental string in Ext$^p(\cS_1, \cS_2)$ is given by
\begin{equation} \label{massformula}
{m_{1 \rightarrow 2}}^2 = \tfrac{1}{2}\,(\xi_2-\xi_1+p-1)\,.
\end{equation}
So, in our case, that statement becomes the following: 

\emph{Near a wall of marginal stability, brane $\cS$ is stable with respect to decay into $\cS_1$ and $\cS_2$ provided $\xi_1 > \xi_2$}.

The full-fledged statement of $\Pi$-stability is essentially the same as what we have just said, except that we may now deal with general objects of the bounded derived category of coherent sheaves on $X_3$, $D^b(X_3)$, as opposed to just coherent sheaves. This also means that the notion of short exact sequence gets replaced by that of \emph{distinguished triangles}. 

\subsection{$\Pi$-stability for $O7/D7$ systems}
Everything in the preceding section was defined to deal with Dp-branes wrapping p-cycles of a CY threefold. However, the full power of that language can be transposed to the setting of CY threefold compactifications with spacetime filling D-branes and O-planes, leading to $d=4, \mathcal{N}=1$ theories. All one needs to do is to add three spatial worldvolume dimensions to the branes. Although the term `central charge' no longer has a clear physical meaning, the formalism still allows one to decide which branes preserve $\mathcal{N}=1$. One must first compute the phase of central charge $\xi_{O7}$ associated to the O7-plane with formula \eqref{centralcharge}, but with an overall minus sign:
\begin{equation}
\xi_{O7} = \frac{1}{\pi}\, {\rm arg}\left(\int_{X_3} e^{-t}\, (-8\,[D_{O7}]+\tfrac{1}{6}\,\chi(D_{O7})\,\omega\, ) \right)\,, 
\qquad \stackrel{\rm large vol}\longrightarrow \xi_{O7}=0\,.
\end{equation}
This sets the $\mathcal{N}=1$ preserved by the O-plane projection, of the original $\mathcal{N}=2$ preserved by the CY threefold. Borrowing the language from the previous section, let us compute the central charge of a D7-brane charge vector:
\begin{equation}
\Gamma = q_{D9} + {q_{D7}}^i \, D_i + {q_{D5}}_i \, C^i + q_{D3}\, \omega
\end{equation}
with a basis $\{D_i\} \in H^2(X_3, \mathbb{Z})$ of $(1,1)$ forms, a dual basis of four-forms $\{C^i\} \in H^4(X_3, \mathbb{Z})$ such that $\int_{X_3} D_i \wedge C^j = {\delta_i}^j$, and $\omega$ the volume form of norm one. Defining the triple intersection numbers 
\be 
c_{i j k} \equiv \int_{X_3} D_i \wedge D_j \wedge D_j\,,
\ee
we can write down the central charge as a function of $t = (B^k+i\,J^k)\,D_k$ as follows
\be Z = -\int_{X_3} \exp(-t) \wedge \Gamma = -\tfrac{1}{2}\,{q_{D7}}^i\,t^j\,t^k\,c_{i j k}+{q_{D5}}_i\,t^i - q_{D3}\,,
\ee
where we took $q_{D9}=0$.
If we take a large volume limit where all ratios\footnote{There may be other, more subtle `large volume limits' one could take. In the case of O7-plane projections with $h^{1,1}_-=0$, the B-field has a fixed value.} $B_k/J_k \rightarrow 0$, this reduces to
\be Z \longrightarrow \tfrac{1}{2}\,{q_{D7}}^i\,J^j\,J^k\,c_{i j k}+i\,{q_{D5}}_i\,J^i - q_{D3}\,.
\ee
Hence, the requirement that a D7-brane preserves the same $\mathcal{N}=1$ boils down to the constraint
\be
\sum_i {q_{D5}}_i\,J^i = 0\,.
\ee
Note that, in these conventions, it is anti-D3-branes that are supersymmetric.

In the next section, we will study the stability condition for a D7-brane of charge vector $\Gamma$ with respect to decay into two D7-branes $\Gamma \rightarrow \Gamma_1 + \Gamma_2$ directly in an example.

\subsection{Examples of T-branes in a one-modulus compactification}
The T-branes we have discussed until now are only to be thought of as local descriptions. Depending on how the models are compactified, these may or may not satisfy the full set of equations of motion. The devil is in the D-terms. In order to have a well-defined, normalizable solution that preserves both Poincar\'e invariance and some fraction of supersymmetry, the simplest possible scenario requires compactifying the 7-branes on a Riemann surface of negative sectional curvature, as discussed in \cite{Anderson:2013rka}.

Compactifying the 7-brane on a simply connected complex surface simplifies matters, as we do not have to deal with Wilson lines in order to define bundles. Hence, in this section, we will present simple examples of globally well-defined T-branes. We will compactify IIB on the octic CY threefold given by a degree eight hypersurface in weighted projective space
\begin{equation}
\P^4_{1, 1, 1, 1, 4}[8] \qquad \text{with homogeneous coordinates} \qquad [x_1: \ldots : x_4: \xi]\,,
\end{equation}
where $\xi$ has degree four. This example is particularly nice because it admits the orientifold involution $\xi \rightarrow -\xi$. The O7-plane is defined by $\xi=0$. Its Poincar\'e dual class is $D_{O7} = 4\,H$, where $H$ is the hyperplane class of the ambient space, defined as the Poincar\'e dual to, say $x_1 = 0$. The triple intersection number of this space is $\int_{X_3} H^3 = 2$\,. There is only one complexified K\"ahler modulus $t H = (B+i J) H$. However, the orientifold projection sets $B=0$, hence our computations for the central charges simplify drastically.

A generic D7-brane will have a charge vector $\Gamma = q_{D7}\,H+ q_{D5}\,(\tfrac{1}{2})\,H^2+ q_{D3}\,(\tfrac{1}{2})\,H^3$ of induced charges. The condition that such a brane be supersymmetric is therefore, that its phase be real. In this case, the large volume limit yields
\begin{equation}
Z(D7) = J^2\,q_{D7}+i\,J\,q_{D5}-q_{D3}\,.
\end{equation}
The requirement, therefore, boils down to having zero induced D5-charge. In this compactification, orientifold-invariant branes already satisfy this, therefore only brane-image-brane pairs can violate it.
This has drastic consequences in this case: \emph{No two D7-branes with a chiral spectrum can be mutually supersymmetric in one-modulus compactifications!} Any such pair will inextricably: a) form a bound state, either as a smooth recombined brane, or as a T-brane; b) or there will be a non-supersymmetric vacuum.

Let us take two generic D7-branes in our octic compactification, and define the combined system as a coherent sheaf as follows:

\be \label{tachyoncompacttbrane}
\begin{tikzcd}[column sep=50pt, ampersand replacement=\&]
0 \rightarrow  \cO(-d_1+f_1) \oplus \cO(-d_2+f_2) \arrow{r}{\bpp P_{1} & 0 \\ 0 & P_{2} \epp} \& \cO(f_1) \oplus \cO(f_2) \rightarrow \cS_1 \oplus \cS_2 \rightarrow 0
\end{tikzcd}\,.
\ee

Concretely, each $\cS_i$ is a D7-brane wrapping a divisor $P_{i}$ of degree $d_i$, carrying worldvolume 
flux $F = (f_i-d_i/2)\,H$. Let us simplify this general discussion by choosing $d_1=d_2=d$, and impose that the net D5-charge of the combined system be zero, which translates to $f_1+f_2=d$. 

Any two such branes will intersect along a curve $\cC$ that will host charged matter. The latter can be computed via Ext groups either directly, or via a spectral sequence. The result is the following for the chiral and anti-chiral matter, respectively:
\begin{eqnarray}
{\rm Ext}^1(\cS_1, \cS_2)) &=& H^0(\cC, \cO(f_2-f_1+d_1)) = H^0(\cC, \cO(2\,f_2))\,,\\
{\rm Ext}^1(\cS_2, \cS_1)) &=& H^0(\cC, \cO(f_1-f_2+d_2)) = H^0(\cC, \cO(2\,f_1))\,. \nonumber
%\qquad \cong {\rm Ext}^2(\cS_1, \cS_2))^* = H^1(\cC, \cO(f_2-f_1+d_1))^* \nonumber
\end{eqnarray}
%where we used Serre duality in the last two expressions. 

By inspecting the tachyon matrix in \eqref{tachyoncompacttbrane}, we see immediately that ${\rm Ext}^1(\cS_1, \cS_2)$ corresponds to the set of possible $(2,1)$ entries modulo automorphisms that kill off anything in the ideal $(P_{1}, P_{2})$, and, similarly, ${\rm Ext}^1(\cS_2, \cS_1)$ corresponds to the $(1,2)$ entries.
These fields, call them, ${\chi_+}^i$ and ${\chi_-}^j$, respectively, must satisfy the following D-term constraint:
\begin{equation} \label{dterm}
\sum_i |{\chi_+}^i|^2-\sum_j |{\chi_-}^j|^2 = \mu (\xi_1-\xi_2) \approx \frac{2\, \mu}{\pi\,J}\,(f_1-f_2)\,,
\end{equation}
where $\mu$ is a positive constant such that, this D-term reproduces the mass formula in \eqref{massformula} correctly for both kinds of fields. Note, that, in principle, vector-like pairs might be lifted by instanton-generated F-terms. We will assume that such effects are absent or suppressed in this discussion. If such effects do occur, however, then one would conclude by inspecting the D-term \eqref{dterm}, that T-branes are impossible in one-modulus Calabi-Yau's.

We will now analyze two special cases of interest:
\paragraph{Case 1: Constant T-brane}
If we choose $f_2=0, f_1=d$, then, the $(2,1)$ entry, the positively charged matter, is a constant (section of $\cO$). The negatively charged matter is a section of $\cO(2\,d)$. The right-hand-side of the D-term \eqref{dterm} is then positive. Here we get a family of solutions, including one with a tachyon $T$ of the form
\begin{equation}
T = \begin{pmatrix}
P_1 & 0 \\ c & P_2
\end{pmatrix} \qquad {\rm with} \quad c = 2\,d/J\,.
\end{equation}
This can be diagonalized to the following system
\be
\begin{tikzcd}[column sep=50pt, ampersand replacement=\&]
\cO(-d) \oplus \cO \arrow{r}{\bpp P_{1} P_2/c & 0 \\ 0 & c \epp} \& \cO(d) \oplus \cO  \qquad \cong \qquad
\cO(-d)  \arrow{r}{ P_{1} P_2/c } \& \cO(d)\,.
\end{tikzcd}
\ee
We manifestly see that the brane behaves as a single brane on a reducible divisor, with the gauge group broken to a single `center of mass $U(1)$'. Because the new brane has zero D5-charge, it preserves the same supersymmetry as the O7-plane. We emphasize that the stability conditions force this brane recombination upon us.

\paragraph{Case 2: Point-like matter}
Let us now choose $f_2>0$ and $f_1>f_2$. In this case, there exists again a family of solutions to the D-term equations, with a special case being that where the $(1,2)$ entry is zero, and the $(2,1)$ entry is a section of $\cO(2\,f_2)$. 
\begin{equation}\label{TbraneRho}
T = \begin{pmatrix}
P_1 & 0 \\ \rho & P_2
\end{pmatrix} \qquad {\rm with} \quad \rho \in \cO_\cC(2\,f_2)\,.
\end{equation}
This is a non-diagonalizable T-brane that has point-like matter of the kind discussed in section \ref{sec:pointlike}, along the points $P_1=P_2=\rho=0$.

The analog of \eqref{AntiD3bound} in this case is
\begin{equation}
\begin{tikzpicture}[scale=1.8]
\node (D711) at (4,2) {$\cO(f_2)$};
\node (D712) at (6,2) {$\underline{\cO(f_2+d)}$};
\node (D7+) at (8,2) { \, D$7_2$};
\node (D31) at (1,1) {$\cO(f_1-2\,d)$};
\node (D32) at (2.5,1) {$E$};
\node (D33) at (4,1) {$\underline{F}$};
\node (D34) at (6,1) {$\cO(f_2+d)$};
\node (D3bar) at (8,1) {$\overline{D3}$};
\node (D721) at (1,0) {$\cO(f_1-2\,d)$};
\node (D722) at (2.5,0) {$\underline{\cO(f_1-d)}$};
\node (D7-) at (8,0) { \, D$7_1$};

\path[->,font=\scriptsize]
(D711) edge node[auto] {$P_2$} (D712)
(D31) edge node[auto] {$-X_2$} (D32)
(D32) edge node[auto] {$-X_1$} (D33)
(D33) edge node[auto] {$-X_0$} (D34)
(D721) edge node[below] {$P_1$} (D722)
(D712) edge[thick, blue] node[black, auto] {$ 1 $}(D34)
(D711) edge[thick, blue] node[black, auto] {$\bp 0 \\ 0 \\ 1 \ep$}(D33)
(D31) edge[thick, red] node[black, auto] {$1$}(D721)
(D32) edge[thick, red] node[black, auto] {$\bp 0 & 0 & 1 \ep$}(D722)
(D7+) edge[blue, thick] (D3bar)
(D3bar) edge[red, thick] (D7-)
;
\end{tikzpicture}
\end{equation}

where
\begin{eqnarray} \nonumber
X_0 &\equiv& \bp \rho &-P_1 & -P_2 \ep\,, \quad X_1 \equiv \bp  P_1 & 0 & P_2  \\  \rho & P_2 & 0 \\  0 & -P_1 & \rho \ep\,, \quad X_2 \equiv \bp  P_2 \\ -\rho \\-P_1 \ep\,,\\ \nonumber \\
{\rm and} \quad E &=& \cO(f_1-d) \oplus \cO(f_2-d) \oplus \cO(f_1-d), \qquad
F = \cO(f_1) \oplus \cO(f_2) \oplus \cO(f_2) \nonumber
\end{eqnarray}
After appropriate automorphisms, this reduces to
\be
\begin{tikzcd}[column sep=50pt, ampersand replacement=\&]
\cO(f_1-d) \oplus \cO(f_2-d) \arrow{r}{\bpp P_{1} & 0 \\ \rho & P_2 \epp} \& \cO(f_1) \oplus \cO(f_2)  
\end{tikzcd}\,.
\ee
Notice that the end result is our T-brane bound state of two D7-branes with fluxes $f_1-d/2$ and $f_2-d/2$, whereas the initial description is as a bound state of an anti-D3 with two D7-branes with fluxes $f_1-\tfrac{3}{2}\,d$ and $f_2+d/2$. So, this T-brane can be disentangled in two inequivalent ways. This is only visible in a compact space, where we have gradings to distinguish these branes.

Let us now study the stability conditions for this T-brane to decay into these three constituents. There are two possible decay processes: 
\begin{enumerate}
\item T-brane $\rightarrow$ D$7_1+\overline{\rm D}3/{\rm D7}_2$ $\rightarrow$ D$7_1+\overline{\rm D}3+{\rm D7}_2$. 
\item T-brane $\rightarrow$ D$7_1/\overline{\rm D}3+{\rm D7}_2$ $\rightarrow$ D$7_1+\overline{\rm D}3+{\rm D7}_2$. 
\end{enumerate}

In either case, the first decay has as the right-hand-side of the D-term constraint $(f_2-f_1+2\,d)$. So, the bound state is stable only if this is positive. Using that $f_1+f_2=d$ for absence of total D5-brane charge, we see that for both process 1. and 2., the first step can take place if $f_2<-d/2$. Let us now discuss the second decay. In process 1., the second decay can take place if $f_2<-d/2$, whereas in process 2., the second decay can take place if $f_1>3d/2$, which again is equivalent to $f_2<-d/2$.

To summarize, assuming absence of total D5-brane charge, the condition of stability of the T-brane \eqref{TbraneRho} against any type of decay is $0<f_2<d/2$. On the other hand, if $f_2>d/2$, the T-brane will decay into the two D7-branes defined by \eqref{tachyoncompacttbrane}. However, the case with $f_2<-d/2$, for which the T-brane would decay into two D7's and one anti-D3 according to either of the processes discussed above, does not exist, since $\rho$ would be a section of a negative bundle, and hence forbidden from assuming a vev. Therefore, whenever this T-brane can exist holomorphically, it is also automatically stable against these two decay channels involving anti-D3's. This result bolsters our claim that T-branes with point-like matter can be thought of as bound states of 7-branes with 3-branes.

\section{Discussion}\label{DiscSec}

In this paper, we have analyzed wide classes of configurations with perturbative IIB D7-branes. The main conclusion is the following:
Whenever the tachyon matrix can be diagonalized, calculations of gauge groups and spectra become extremely simple. Whenever the off-diagonal entries are not constant, hence preventing such diagonalizations, matter fields exhibit strange behavior regarding their localization. In all such cases, we find that there is a lower-dimensional brane lurking inside the 7-brane underlying the observed localization. 

Although our techniques are manifestly perturbative in nature, one cannot help but speculate that more general F-theory T-branes should also obey this simple principle. Instead of only D5 and D3-branes, one might expect to also have $(p, q)$ 5-branes around.
From the dual M-theory picture, one would expect to see, along side with M2-branes, M5-branes that `dissolve' into the background. 

The tachyon condensation picture is the most unified framework to describe perturbative IIB D-branes. To date, there is no SL$(2, \mathbb{Z})$-covariant framework that generalizes it, in order to describe, say, $(p, q)$ 5-branes or 7-branes. Therefore, although we believe that this paper simplifies and elucidates the peculiar behavior of perturbative T-branes, F-theory T-branes will require a new approach. We find the question worth pursuing, and will propose some ideas on how to go about it in our follow up paper \cite{Collinucci:2014taa}. What we will show there, is that one can describe the F-theory lift of certain T-branes by placing coherent sheaves \emph{on the F-theory CY fourfold}. This has no straightforward physical interpretation, as F-theory is not a theory of branes with strings attached. Nevertheless, we will find evidence that such an approach can still capture non-trivial information about the spectrum of F-theory on singular spaces.

\section*{Acknowledgements}
We would like to thank Riccardo Argurio, Andreas Braun, Pierre Corvilain, Marco Fazzi, Hirotaka Hayashi, Ruben Minasian and Roberto Valandro for useful and stimulating discussions.

A. C. is a Research Associate of the Fonds de la Recherche Scientifique F.N.R.S. (Belgium).
The work of R.S. was supported by the ERC Starting Independent Researcher Grant 259133-ObservableString.

R.S. would like to thank the Universit\'e Libre de Bruxelles for hospitality during part of this project.
The authors are grateful to the Mainz Institute for Theoretical Physics (MITP) for its hospitality and its partial support during the completion of this work.

\bibliography{../../u1.bib}

\providecommand{\href}[2]{#2}\begingroup\raggedright\begin{thebibliography}{10}

\bibitem{Donagi:2003hh}
R.~Donagi, S.~Katz, and E.~Sharpe, ``{Spectra of D-branes with higgs vevs},''
  \href{http://dx.doi.org/10.4310/ATMP.2004.v8.n5.a3}{{\em
  Adv.Theor.Math.Phys.} {\bf 8} (2005)  813--859},
\href{http://arxiv.org/abs/hep-th/0309270}{{\tt arXiv:hep-th/0309270
  [hep-th]}}.
%%CITATION = HEP-TH/0309270;%%.

\bibitem{Donagi:2011jy}
R.~Donagi and M.~Wijnholt, ``{Gluing Branes, I},''
  \href{http://dx.doi.org/10.1007/JHEP05(2013)068}{{\em JHEP} {\bf 1305} (2013)
   068},
\href{http://arxiv.org/abs/1104.2610}{{\tt arXiv:1104.2610 [hep-th]}}.
%%CITATION = ARXIV:1104.2610;%%.

\bibitem{Cecotti:2010bp}
S.~Cecotti, C.~Cordova, J.~J. Heckman, and C.~Vafa, ``{T-Branes and
  Monodromy},'' \href{http://dx.doi.org/10.1007/JHEP07(2011)030}{{\em JHEP}
  {\bf 1107} (2011)  030},
\href{http://arxiv.org/abs/1010.5780}{{\tt arXiv:1010.5780 [hep-th]}}.
%%CITATION = ARXIV:1010.5780;%%.

\bibitem{Anderson:2013rka}
L.~B. Anderson, J.~J. Heckman, and S.~Katz, ``{T-Branes and Geometry},''
  \href{http://dx.doi.org/10.1007/JHEP05(2014)080}{{\em JHEP} {\bf 1405} (2014)
   080},
\href{http://arxiv.org/abs/1310.1931}{{\tt arXiv:1310.1931 [hep-th]}}.
%%CITATION = ARXIV:1310.1931;%%.

\bibitem{Sen:1998sm}
A.~Sen, ``{Tachyon condensation on the brane anti-brane system},''
  \href{http://dx.doi.org/10.1088/1126-6708/1998/08/012}{{\em JHEP} {\bf 9808}
  (1998)  012},
\href{http://arxiv.org/abs/hep-th/9805170}{{\tt arXiv:hep-th/9805170
  [hep-th]}}.
%%CITATION = HEP-TH/9805170;%%.

\bibitem{Douglas:2000gi}
M.~R. Douglas, ``{D-branes, categories and N=1 supersymmetry},''
  \href{http://dx.doi.org/10.1063/1.1374448}{{\em J.Math.Phys.} {\bf 42} (2001)
   2818--2843},
\href{http://arxiv.org/abs/hep-th/0011017}{{\tt arXiv:hep-th/0011017
  [hep-th]}}.
%%CITATION = HEP-TH/0011017;%%.

\bibitem{Freed:1999vc}
D.~S. Freed and E.~Witten, ``{Anomalies in string theory with D-branes},'' {\em
  Asian J.Math} {\bf 3} (1999)  819,
\href{http://arxiv.org/abs/hep-th/9907189}{{\tt arXiv:hep-th/9907189
  [hep-th]}}.
%%CITATION = HEP-TH/9907189;%%.

\bibitem{Witten:1998cd}
E.~Witten, ``{D-branes and K theory},''
  \href{http://dx.doi.org/10.1088/1126-6708/1998/12/019}{{\em JHEP} {\bf 9812}
  (1998)  019},
\href{http://arxiv.org/abs/hep-th/9810188}{{\tt arXiv:hep-th/9810188
  [hep-th]}}.
%%CITATION = HEP-TH/9810188;%%.

\bibitem{Minasian:1997mm}
R.~Minasian and G.~W. Moore, ``{K theory and Ramond-Ramond charge},''
  \href{http://dx.doi.org/10.1088/1126-6708/1997/11/002}{{\em JHEP} {\bf 9711}
  (1997)  002},
\href{http://arxiv.org/abs/hep-th/9710230}{{\tt arXiv:hep-th/9710230
  [hep-th]}}.
%%CITATION = HEP-TH/9710230;%%.

\bibitem{Sharpe:1999qz}
E.~R. Sharpe, ``{D-branes, derived categories, and Grothendieck groups},''
  \href{http://dx.doi.org/10.1016/S0550-3213(99)00535-0}{{\em Nucl.Phys.} {\bf
  B561} (1999)  433--450},
\href{http://arxiv.org/abs/hep-th/9902116}{{\tt arXiv:hep-th/9902116
  [hep-th]}}.
%%CITATION = HEP-TH/9902116;%%.

\bibitem{Aspinwall:2004jr}
P.~S. Aspinwall, ``{D-branes on Calabi-Yau manifolds},''
\href{http://arxiv.org/abs/hep-th/0403166}{{\tt arXiv:hep-th/0403166
  [hep-th]}}.
%%CITATION = HEP-TH/0403166;%%.

\bibitem{Sharpe:2003dr}
E.~Sharpe, ``{Lectures on D-branes and sheaves},''
\href{http://arxiv.org/abs/hep-th/0307245}{{\tt arXiv:hep-th/0307245
  [hep-th]}}.
%%CITATION = HEP-TH/0307245;%%.

\bibitem{Collinucci:2014taa}
A.~Collinucci and R.~Savelli, ``{F-theory on singular spaces},''
\href{http://arxiv.org/abs/1410.4867}{{\tt arXiv:1410.4867 [hep-th]}}.
%%CITATION = ARXIV:1410.4867;%%.

\bibitem{Gomez:2000zm}
T.~Gomez and E.~R. Sharpe, ``{D-branes and scheme theory},''
\href{http://arxiv.org/abs/hep-th/0008150}{{\tt arXiv:hep-th/0008150
  [hep-th]}}.
%%CITATION = HEP-TH/0008150;%%.

\bibitem{Heckman:2010qv}
J.~J. Heckman, Y.~Tachikawa, C.~Vafa, and B.~Wecht, ``{N = 1 SCFTs from Brane
  Monodromy},'' \href{http://dx.doi.org/10.1007/JHEP11(2010)132}{{\em JHEP}
  {\bf 1011} (2010)  132},
\href{http://arxiv.org/abs/1009.0017}{{\tt arXiv:1009.0017 [hep-th]}}.
%%CITATION = ARXIV:1009.0017;%%.

\bibitem{Diaconescu:1998ua}
D.-E. Diaconescu and S.~Gukov, ``{Three-dimensional N=2 gauge theories and
  degenerations of Calabi-Yau four folds},''
  \href{http://dx.doi.org/10.1016/S0550-3213(98)00597-5}{{\em Nucl.Phys.} {\bf
  B535} (1998)  171--196},
\href{http://arxiv.org/abs/hep-th/9804059}{{\tt arXiv:hep-th/9804059
  [hep-th]}}.
%%CITATION = HEP-TH/9804059;%%.

\bibitem{Grimm:2011fx}
T.~W. Grimm and H.~Hayashi, ``{F-theory fluxes, Chirality and Chern-Simons
  theories},'' \href{http://dx.doi.org/10.1007/JHEP03(2012)027}{{\em JHEP} {\bf
  1203} (2012)  027},
\href{http://arxiv.org/abs/1111.1232}{{\tt arXiv:1111.1232 [hep-th]}}.
%%CITATION = ARXIV:1111.1232;%%.

\bibitem{Cvetic:2012xn}
M.~Cvetic, T.~W. Grimm, and D.~Klevers, ``{Anomaly Cancellation And Abelian
  Gauge Symmetries In F-theory},''
  \href{http://dx.doi.org/10.1007/JHEP02(2013)101}{{\em JHEP} {\bf 1302} (2013)
   101},
\href{http://arxiv.org/abs/1210.6034}{{\tt arXiv:1210.6034 [hep-th]}}.
%%CITATION = ARXIV:1210.6034;%%.

\bibitem{Hayashi:2013lra}
H.~Hayashi, C.~Lawrie, and S.~Schafer-Nameki, ``{Phases, Flops and F-theory:
  SU(5) Gauge Theories},''
  \href{http://dx.doi.org/10.1007/JHEP10(2013)046}{{\em JHEP} {\bf 1310} (2013)
   046},
\href{http://arxiv.org/abs/1304.1678}{{\tt arXiv:1304.1678 [hep-th]}}.
%%CITATION = ARXIV:1304.1678;%%.

\bibitem{Collingwood}
D.~H. Collingwood and W.~M. McGovern, {\em Nilpotent orbits in semisimple Lie
  algebras}.
\newblock Van Nostrand, 1993.

\bibitem{Douglas:2000ah}
M.~R. Douglas, B.~Fiol, and C.~Romelsberger, ``{Stability and BPS branes},''
  \href{http://dx.doi.org/10.1088/1126-6708/2005/09/006}{{\em JHEP} {\bf 09}
  (2005)  006},
\href{http://arxiv.org/abs/hep-th/0002037}{{\tt arXiv:hep-th/0002037
  [hep-th]}}.
%%CITATION = HEP-TH/0002037;%%.

\end{thebibliography}\endgroup
\bibliographystyle{utphys}

\end{document}